\documentclass[aps,pre,preprint,superscriptaddress]{revtex4-1}
\usepackage{latexsym}
\usepackage{natbib}
\usepackage{amsmath}
\usepackage{amsfonts}
\usepackage{amssymb}
\usepackage[utf8]{inputenc}
\usepackage[T1]{fontenc}
\usepackage[dvips]{color,graphicx}
\usepackage{verbatim}
\usepackage{t1enc}
\usepackage{anysize}
\usepackage{tocbibind}
\usepackage[percent]{overpic}
\usepackage{float} 
\usepackage{appendix}
\usepackage{booktabs} 
\usepackage{natbib}
\usepackage{array}
\usepackage{xcolor}
\usepackage[normalem]{ulem}

\begin{document}
\title{Superexponential growth of epidemics in networks with cliques}

\author{L.  D. Valdez}

\affiliation{Departamento de F\'isica, FCEyN, Universidad Nacional de Mar del Plata, Mar del Plata 7600, Argentina.}
\affiliation{Instituto de Investigaciones F\'isicas de Mar del Plata (IFIMAR), CONICET, Mar del Plata 7600, Argentina.}
  \date{\today}

\begin{abstract}
Many dynamic processes on complex networks, from disease outbreaks to cascading failures, can rapidly accelerate once a critical threshold is exceeded, potentially leading to severe social and economic costs.  Therefore, in order to develop effective mitigation strategies, it is essential to understand how these catastrophic events occur. In this work, we investigate the dynamic of disease propagation on networks with fully connected sub-graphs (or cliques) using a susceptible-infected-quarantined (SIQ) model, and considering a scenario in which only a proportion $f$ of the population has access to testing. For this model, we derive the time-evolution equations governing the spread of epidemics and show that the final proportion of infected individuals undergoes a sudden transition at a critical threshold $f_c$. Moreover, close to this transition point, our results on the time evolution of the SIQ model reveal that the number of new cases can exhibit a faster-than-exponential growth. This accelerated spread dynamics is more likely to occur in networks with larger cliques.\end{abstract}

\maketitle

\section{Introduction}
A wide range of real systems in nature and society often display complex and non-linear dynamics, and under specific conditions, they may display abrupt transitions between different states~\cite{artime2024robustness,kuehn2021universal}. These explosive events can result in significant social and economic consequences, and it may be difficult or even impossible to restore these systems to their original state~\cite{schneider2004abrupt,perrings2009irreversibility}. Therefore, there has been a growing interest in the development of mathematical and computational models to investigate the underlying mechanisms driving these abrupt transitions, in order to predict and prevent such catastrophic events~\cite{artime2024robustness,kuehn2021universal,d2019explosive,d2019explosive,liang2018feedback,xue2024nucleation}.

For example, in the last decade, the study of interdependent networks has emerged as a powerful framework for understanding how interdependent infrastructures respond to perturbations~\cite{gao2022introduction,valdez2020cascading,dong2021briefing,buldyrev2010catastrophic,son2012percolation,li2021percolation}. By using both stochastic simulations and percolation theory, researchers have shown that even small damages in interdependent networks can initiate a cascading failure process that eventually destroys the entire system. Interestingly, near the collapse point, it was found that the failure propagation dynamic is characterized by a long-lasting plateau~\cite{buldyrev2010catastrophic,zhou2014simultaneous}, which could potentially serve as an early warning signal and provide a window of opportunity for targeted and microscopic interventions to halt the cascade.

Similarly, research on the modeling of infectious diseases has shown that social networks are also susceptible to abrupt epidemic transitions~\cite{valdez2024explosive,pires2023tricritical,cui2018epidemic,lamata2024pathways,borner2022explosive,st2021universal,hebert2015complex}. For example, several works revealed that certain social distancing strategies, while intended to reduce the risk of infection, can paradoxically lead to a sudden increase in the final fraction of infected people~\cite{gross2006epidemic,berner2023adaptive}. On the other hand, researchers have found that limited resources for disease control can, as well, induce an abrupt epidemic transition~\cite{scarselli2021discontinuous,lamata2023collapse,di2018multiple,bottcher2015disease}. Moreover, Scarselli et al.~\cite{scarselli2021discontinuous} showed that in a scenario where testing resources are limited, the time evolution of the number of infected people could increase at a faster-than-exponential rate. This phenomenon, known as super-exponential growth~\cite{viboud2016generalized,leiss2015super} has been empirically observed in COVID-19 and influenza data~\cite{wu2021generalized,pavithran2022extreme,scarpino2016effect}. As a result of this accelerated growth, the doubling-time, which is a widely used indicator in epidemiology~\cite{boslaugh2007encyclopedia,anzai2022doubling}, could decrease significantly over time in this "explosive" scenario. In another work, Bulchandani et al.~\cite{bulchandani2021digital} also found that a digital contact tracing strategy (which is a non-pharmaceutical intervention that gained popularity during the COVID-19 pandemic) can also induce an abrupt epidemic transition. More specifically, they studied a strategy in which a fraction $\phi$ of individuals use a contact-tracing app. When one of these app users exhibits symptoms, they are immediately quarantined and a notification is sent to their contacts who also use the app. Those contacts, in turn, notify their own app-using contacts, triggering a recursive process in which all notified individuals are quarantined without delay. For this model, it was also considered that a fraction $\theta$ of the population does not show symptoms. Bulchandani et al.~\cite{bulchandani2021digital} found that for a perfect contact tracing scenario ($\phi=1$), their model exhibits a discontinuous transition at $\theta_c=1$.

In a recent study, Valdez et al.~\cite{valdez2023epidemic} found that an epidemic spreading process can exhibit properties of both continuous and discontinuous phase transitions on networks with cliques (i.e., fully connected sub-graphs representing, for example, households or workplaces). Specifically, they studied a discrete-time susceptible-infected-recovered-quarantined (SIRQ) model on random networks with cliques in which, at each time step: 1) infected people transmit the disease to their susceptible neighbors with probability $\beta$, and ii) infected people are detected and placed in quarantine (along with their neighbors) with probability $f$. On one hand, by using the generating-function technique, they showed that the probability of epidemics vanishes continuously at a critical point $f=f_c$. Additionally, around this critical threshold, it was found that the distribution of the final number of infected people decays as a power-law function, which is a behavior commonly observed in second-order phase transitions~\cite{stauffer2018introduction}. Stochastic simulations, however, revealed that epidemic events are abruptly suppressed around $f_c$ as in a first-order phase transition. While these findings suggest a hybrid nature of the phase transition in the SIRQ model, it is important to note that they were primarily obtained through stochastic simulations, without a rigorous theoretical framework to support them. Moreover, it was not possible to accurately investigate how epidemics behave near the critical point, because the probability of these events  tends to zero, as shown in Ref.~\cite{valdez2023epidemic}.  For this reason, it is necessary to develop equations that allow for a more precise analysis of epidemics near the transition point.  In addition, the time evolution of disease spread was not explored in detail in Ref.~\cite{valdez2023epidemic}. Such temporal analysis, however, could potentially reveal early warning signals indicating that the system is approaching an abrupt transition.

To study the time evolution of epidemic spread in networks with cliques, in the present work, we will explore an SIQ model (with $\beta=1$) which is a simplified version of the SIRQ model studied in Ref.~\cite{valdez2023epidemic}. For this particular case, we will be able to write precise equations that describe the behavior of epidemics in the limit of large network sizes (which is also referred to as the "thermodynamic limit"). Our theoretical equations indicate that an abrupt transition in the final fraction of infected people $I_{tot}$ occurs at a threshold $f_c$. In addition, as we approach this critical threshold, our results show that $I_{tot}$ decreases linearly with $f$, which indicates that this is not a hybrid transition—as hybrid transitions typically exhibit power-law decay— but rather resembles a "Type II" transition (following the classification of abrupt transitions proposed in Ref.~\cite{d2019explosive}). Furthermore, we find that the time evolution of the number of new cases $I_{new}(t)$ starts growing faster than an exponential function, especially for networks containing larger cliques. Notably, this super-exponential growth becomes particularly pronounced near the point of abrupt transition, suggesting that it could serve as an indicator that the system is approaching a transition point.

This paper is organized as follows: Sec.~\ref{sec.modeltheor} introduces our model. In Sec.~\ref{sec.class}, we present the main equations governing the time evolution of our SIQ model. Results are discussed in Sec.~\ref{sec.result}. Finally, our conclusions are given in the last section.

\section{Model}\label{sec.modeltheor}
\subsection{Networks with cliques: model and theory}\label{sec.netmod}
\begin{figure}[ht]
\begin{center}
\begin{overpic}[scale=1]{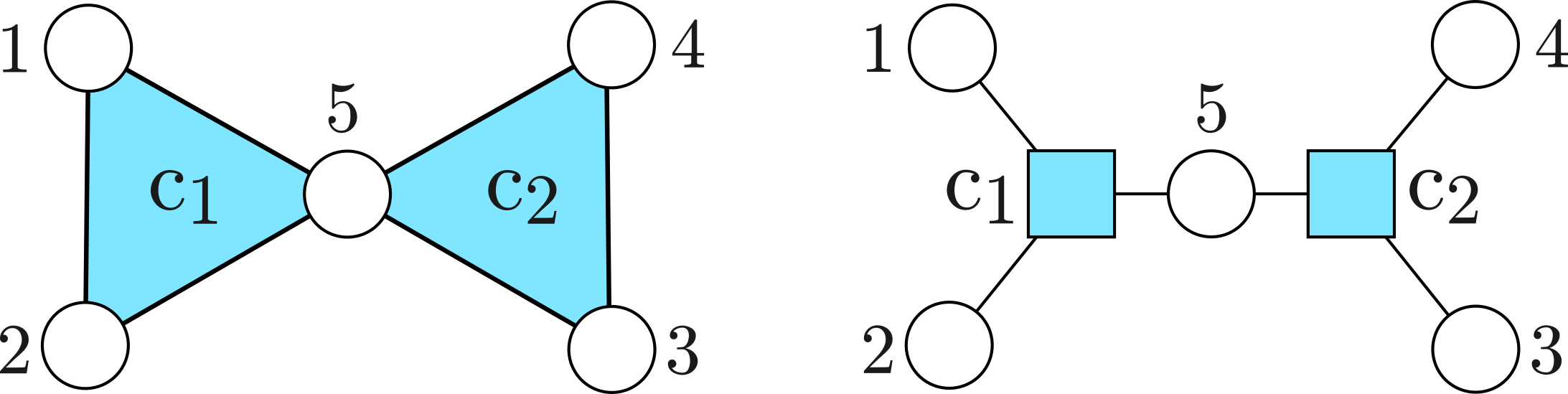}
  \put(-5,13){(a)}
  \put(50,13){(b)}
\end{overpic}
\vspace{-0.5cm}
\end{center}
\caption{Panel (a) illustrates a schematic network with two cliques ($c_1$ and $c_2$) and five individuals, where each clique contains three members. The individuals are labeled with numbers from 1 to 5. Panel (b) displays the bipartite representation of the network from panel (a), where squares represent factor nodes and circles represent individuals.}\label{fig.cliq}
\end{figure}

In this work, we will study the dynamics of disease propagation on complex networks containing fully connected sub-graphs or cliques. As previously noted in Refs.~\cite{mann2023belief,bianconi2024nature,li2021long,karrer2010random}, factor graphs provide a suitable representation for these networks. A factor graph is an undirected bipartite network formed by three sets: nodes (or individuals), factor nodes (or group nodes), and links that exclusively connect nodes with factor nodes. This concept is illustrated in Fig.~\ref{fig.cliq}. The left panel displays a network with cliques, while the right panel shows its associated factor graph. Alternatively, the network on the left can be interpreted as a projection of the factor graph on the right. The degree (or membership) of a node in the factor graph, $k_I$, corresponds to the number of links attached to it, while the degree of a factor node, $k_C$, is defined analogously. The number of nodes and factor nodes are denoted by $N_I$ and $N_C$, respectively. On the other hand, the degree distributions of nodes and factor nodes are represented by $P(k_I)$ and $P(k_C)$, respectively. 

Several methods have been proposed to generate ensembles of random networks with cliques (or factor graphs)~\cite{alves2024clustering,nikolaev2023modeling,leung2016conflicting,guillaume2006bipartite,rizi2024effectiveness}. In this work, we will concentrate on the configuration model~\cite{guillaume2006bipartite,rizi2024effectiveness} which has been widely used to construct random and uncorrelated graphs with a prescribed degree distribution. As it was pointed out in Ref.~\cite{karrer2010random}, sparse graphs generated by the configuration model exhibit a locally tree-like structure in the thermodynamic limit, and thanks to this property, many statistical metrics characterizing the structure of these networks can be calculated by using the generating-function technique. For factor graphs, the following two pairs of probability-generating functions (pgf's) for nodes and factor nodes are commonly employed~\cite{newman2001random}:
\begin{eqnarray}
G_{0}(x)&=&\sum_{k_I=0}^{\infty} P(k_I) x^{k_I}, G_{1}(x)=\sum_{k_I=1}^{\infty} \frac{k_I P(k_I)}{\langle k_I\rangle}x^{k_I-1},\label{eq.g01cG} \\
F_{0}(x)&=&\sum_{k_C=0}^{\infty} P(k_C) x^{k_C}, F_{1}(x)=\sum_{k_C=1}^{\infty} \frac{k_C P(k_C)}{\langle k_C\rangle}x^{k_C-1},\label{eq.g01c}
\end{eqnarray}
where $x$ is a dummy variable, and:
\begin{enumerate}
\item  $\langle k_I\rangle=\sum k_I P(k_I)$ and $\langle k_C\rangle=\sum k_C P(k_C)$ are the expected degree of nodes and factor nodes, respectively.
\item $G_{0}(x)$ is the pgf for the degree of a randomly chosen node/person in the factor graph.
\item $G_{1}(x)$ is the generating function for the number of additional links a node has when it is reached by following a random link in the factor graph. This number is commonly referred to as the excess-degree.
\item $F_{0}(x)$ is the pgf for the degree of a randomly chosen factor node.
\item $F_{1}(x)$ is the generating function for the number of additional links a factor node has when it is reached by following a random link in the factor graph. 
\end{enumerate}
To make these concepts more concrete, consider a scientific collaboration network represented as a bipartite graph where authors/scientists (nodes) connect to their publications (factor nodes). In this example, $G_{0}(x)$ is the generating function for the probability of randomly selecting a scientist who has contributed to a number $k_I$ of publications. On the other hand, $F_0(x)$ is the generating function for the probability of randomly selecting a publication written by $k_C$ scientists.

Thanks to the locally tree-like character of these generated factor graphs, we can also calculate the degree distribution of their associated projected networks using the generating
functions given in Eqs.~(\ref{eq.g01cG}) and~(\ref{eq.g01c}). More specifically, Newman et al~\cite{newman2001random} showed that, by using the "power property of generating functions", the probability generating function for the degree distribution of the projection can be expressed as:
\begin{eqnarray}\label{eq.gvec1}
\mathcal{G}^{(1)}_{0}(x_1) = G_{0}(F_{1}(x_1))=\sum_{n=0}^{\infty}a_n (x_1)^n,
\end{eqnarray}
where $x_1$ is a dummy variable and $a_n=\frac{1}{n!}\frac{d^n \mathcal{G}^{(1)}_{0}(x_1)}{d(x_1)^n}|_{x_1=0}$ represents the probability that a randomly chosen node has $n$ first-nearest neighbors in the projected network. Returning to our scientific collaboration example, $\mathcal{G}^{(1)}_{0}(x_1)$ generates the distribution of the total number of unique collaborators for a randomly selected scientist.

\begin{figure}[ht]
\begin{center}
\begin{overpic}[scale=0.6]{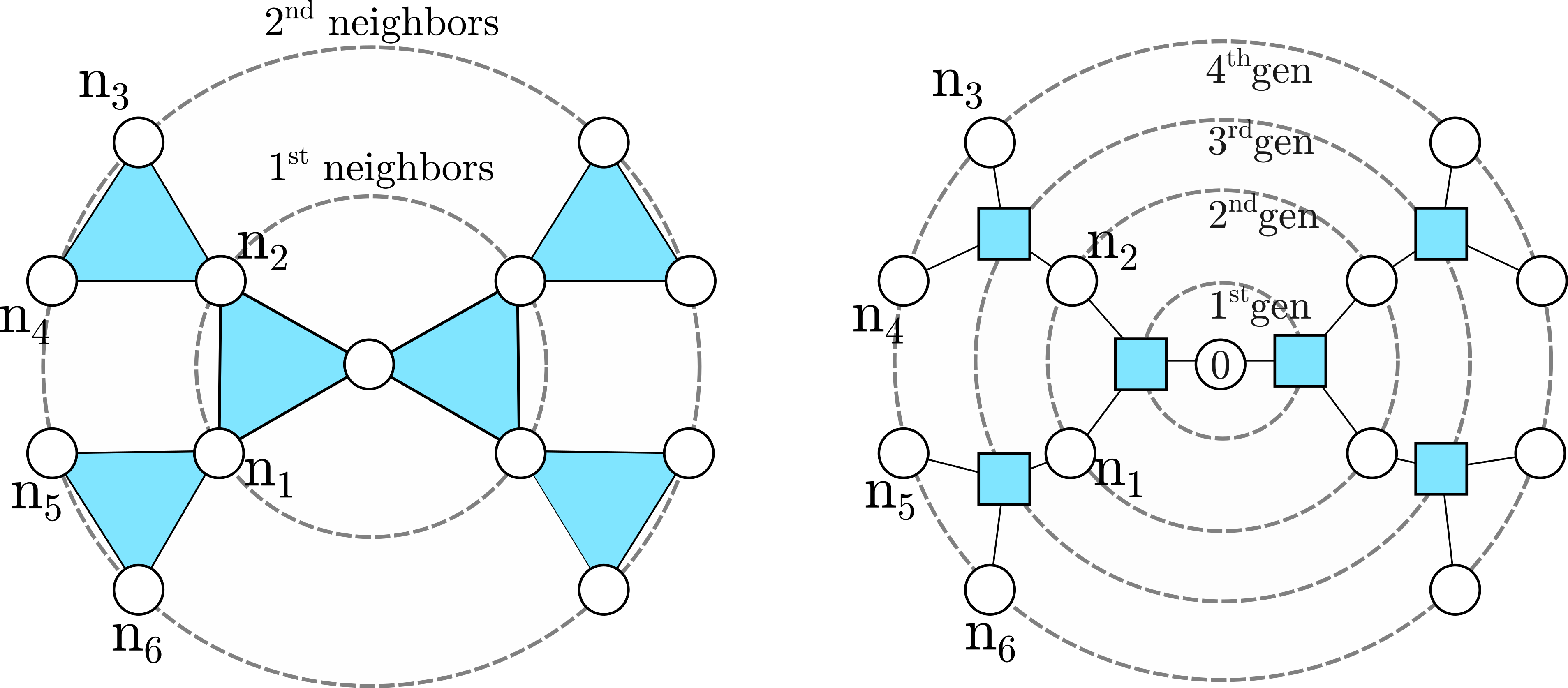}
  \put(85,18){}
\end{overpic}
\vspace{-0.5cm}
\end{center}
\caption{This schematic illustrates how a network with a tree-like structure can be viewed as a branching process. On the left side, we show a simple network with cliques where the central node has four first-neighbors and eight second-neighbors. For instance, $n_1$ and $n_2$  are first-neighbors of the central node, while $n_3$, $n_4$, $n_5$, and $n_6$ are examples of second-neighbors. On the right side, we present the factor graph representation of the network shown on the left, which can be described as a branching process. In this process, the central node (generation 0) acts as the ancestor, from which factor nodes (generation 1) emerge. The total number of these factor nodes is determined by the generating function
$G_{0}(x)$. These factor nodes subsequently have descendants that constitute generation 2 and correspond to the first-neighbors shown on the left panel. The number of individuals produced by \textbf{each} factor node (in the branching process) is described by the generating function $F_{1}(x)$ (see Eq.~(\ref{eq.g01c})). On the other hand, the generating function for the \textbf{total} number of descendants in generation 2 is $G_{0}(F_{1}(x))\equiv \mathcal{G}^{(1)}_0(x)$, which is obtained by using the power property of generating functions~\cite{newman2001random}. By repeating this branching process, we have that the total number of descendants in generation 3 (composed of factor nodes) is given by $G_{0}(F_{1}(G_{1}(x)))$ and for generation 4 (composed of nodes), is given by $G_{0}(F_{0}(G_{1}(F_{1}(x))))\equiv \mathcal{G}^{(2)}_0(x)$. The latter also corresponds to the generating function for the total number of second-neighbors of the central node.}\label{fig.branc}
\end{figure}

Alternatively, the pgf $\mathcal{G}^{(1)}_{0}(x_1)$ given in Eq.~(\ref{eq.gvec1}) can also be understood from a branching process perspective.  Imagine a randomly chosen node as the ancestor or root from which lineages branch off (see Fig.\ref{fig.branc}). The factor nodes connected to this ancestor form the first generation and their number is captured by the outer function $G_0(\cdot)$ in Eq.~(\ref{eq.gvec1}). Each factor node then produces a number of "descendants" or nodes that form the second generation, and this number is encoded in the inner function $F_{1}(\cdot)$ in Eq.~(\ref{eq.gvec1}).

Another distribution that describes the local structure of projected networks, which can be easily computed using the method presented in Ref.~\cite{newman2001random}, is the excess-degree distribution~\footnote{Here, the term 'excess-degree' refers to the number of outgoing first-neighbors of a node that is selected in a two-step process:  first, we randomly choose a clique in the projected network, and then we randomly choose a node inside this clique.}. Newman et al~\cite{newman2001random} showed that the pgf for the excess-degree distribution in these networks is given by:
\begin{eqnarray}\label{eq.gvec1prim}
   \mathcal{G}_{1}^{(1)}(x_1) = G_{1}(F_{1}(x_1)). 
\end{eqnarray}

\begin{figure}[ht]
\begin{center}
\begin{overpic}[scale=0.8]{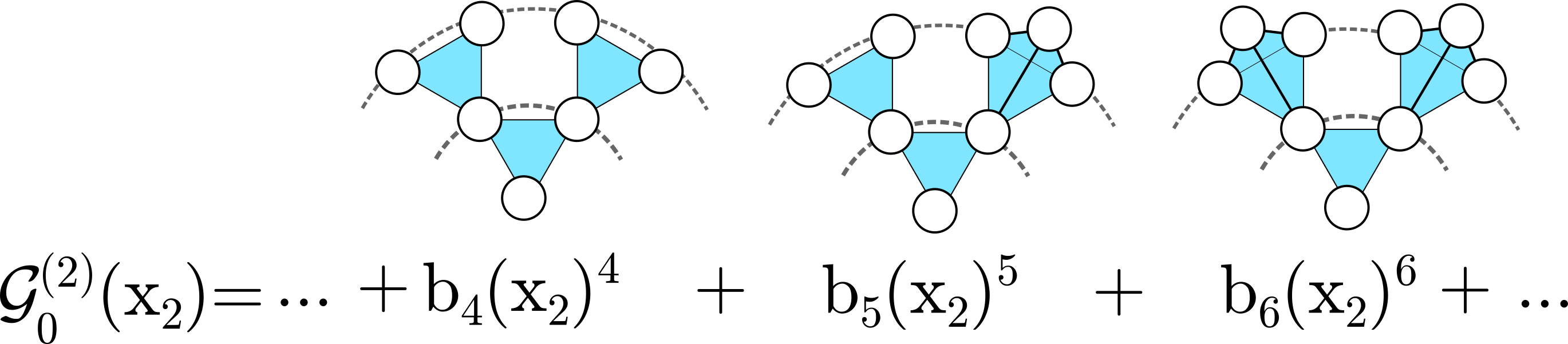}
  \put(85,18){}
\end{overpic}
\end{center}
\vspace{-0.5cm}
\caption{A graphical representation of $\mathcal{G}^{(2)}_0(x_2)$.}\label{fig.g2bip}
\end{figure}

By applying the same approach previously discussed, we can also derive the pgf's for higher-order neighbors in projected networks. For instance, the pgf for the number of second-nearest neighbors of a randomly chosen individual is:
\begin{eqnarray}\label{eq.gvec2}
    \mathcal{G}^{(2)}_0(x_2) &=& G_{0}(F_{1}(G_{1}(F_{1}(x_2)))),\nonumber\\
    &=&\mathcal{G}_0^{(1)}(\mathcal{G}_1^{(1)}(x_2))=\sum_{n=0}^{\infty}b_n (x_2)^n,
\end{eqnarray}
where $x_2$ serves as a placeholder, and $b_n=\frac{1}{n!}\frac{d^n \mathcal{G}^{(2)}_{0}(x_2)}{d(x_2)^n}|_{x_2=0}$ is the probability that the chosen node has $n$ second-nearest neighbors. In Fig.~\ref{fig.g2bip}, we show a graphical representation of $\mathcal{G}^{(2)}_0(x_2)$. This pgf can also be understood in terms of a branching process (as shown in Fig.\ref{fig.branc}): the outer function $\mathcal{G}_0^{(1)}(\cdot)$ in Eq.~(\ref{eq.gvec2}) corresponds to the first-neighbors of a randomly chosen node, and the inner function  $\mathcal{G}_1^{(1)}(\cdot)$ corresponds to the first-neighbors of those first-neighbors. 

Similarly, if we randomly choose an individual through a clique, the pgf for the number of outgoing second-neighbors is:
\begin{eqnarray}\label{eq.gvec2out}
    \mathcal{G}^{(2)}_{1}(x_2) &=& G_{1}(F_{1}(G_{1}(F_{1}(x_2)))),\\
    &=&\mathcal{G}_1^{(1)}(\mathcal{G}_1^{(1)}(x_2)).
\end{eqnarray}

To give a concrete example of the pgf's defined in this section, consider a random network where every person belongs to exactly two cliques, and every clique contains three members. For this case, we have the following generating functions: $G_{0}(x)=x^2$, $G_{1}(x)=x$, $F_{0}(x)=x^3$,  $F_{1}(x)=x^2$. Then, by combining these functions, we get the following:
\begin{itemize} 
\item $\mathcal{G}^{(1)}_{0}(x_1) = G_{0}(F_{1}(x_1)) =(x_1)^4$, which indicates that each person has four neighbors in the projected network.
\item $\mathcal{G}^{(2)}_0(x_2) = G_{0}(F_{1}(G_{1}(F_{1}(x_2)))) =(x_2)^8$, which indicates that each person has eight second-neighbors in the projected network.
\end{itemize}

\subsection{Epidemic modeling}\label{sec.epimodl}
On top of the networks described earlier, we will investigate a compartmental susceptible-infected-quarantined (SIQ) model  in which individuals who test positive are placed in quarantine (along with their neighbors). In our model, susceptible people are healthy but at risk of infection, infected people can transmit the disease to susceptible neighbors, and quarantined people are no longer in contact with the rest of the population and cannot transmit the disease. For this model, we redefine the parameter $f$ presented in Ref.~\cite{valdez2023epidemic} to denote the proportion of the population who have regular access to testing. More specifically, in our model, we differentiate between two types of individuals:
\begin{itemize}
    \item Untested people $(1-f)$: those without access to testing.
    \item Tested people ($f$): those who have regular access to fast and accurate testing. If they test positive, they immediately quarantine themselves and then impose quarantine on their neighbors to prevent further spread.
\end{itemize}
For simplicity, we will assume that the people with and without access to testing
are randomly distributed across the network nodes. On the other hand, we define $S_{\ell}$ and $S_{r}$ as the fractions of susceptible people with and without access to testing, respectively.

\subsubsection{Dynamics of the model}
\begin{figure}[ht]
\begin{center}
\begin{overpic}[scale=0.7]{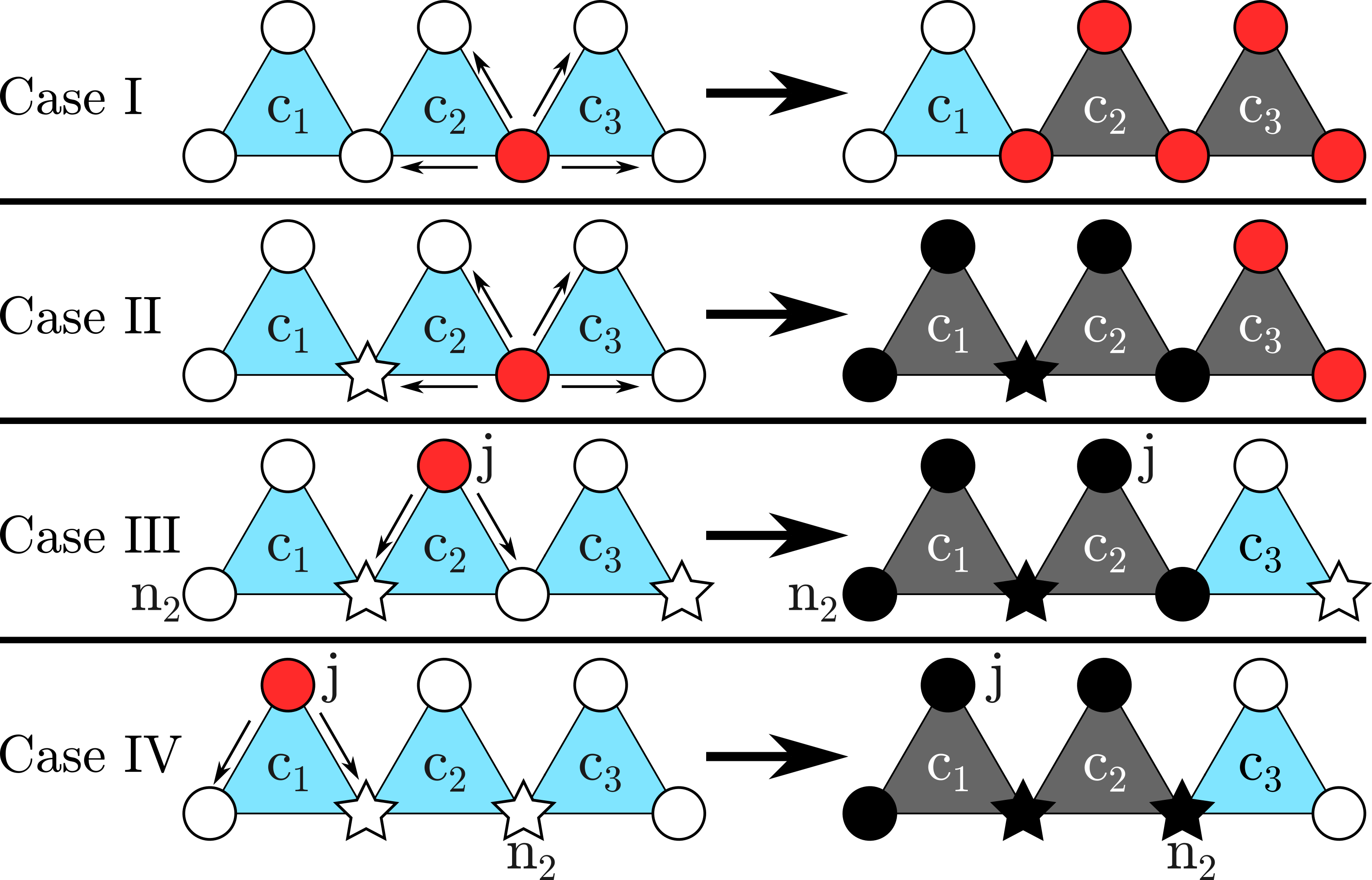}
  \put(85,18){}
\end{overpic}
\vspace{0cm}
\end{center}
\caption{This figure illustrates various scenarios showing how the states of individuals evolve from time $t$ to time $t+1$ in our SIQ model. Stars represent people who are regularly tested and circles represent people who do not have access to testing services. Nodes are color-coded as follows: red for infected, white for susceptible, and black for quarantined. Thin arrows indicate the direction of disease transmission. Each case in the figure displays a configuration at time $t$ on the left panel and the corresponding state transitions at time $t+1$ on the right. Cliques are labeled as $c_1$, $c_2$, and $c_3$ in the center. "Open cliques" are colored in light-blue, and "closed cliques" in dark gray (see Sec.~\ref{sec.class}). Notice that individuals can influence not only their first-nearest neighbors but also their second-nearest neighbors within a single time step. This "second-order influence" is illustrated, for example, in Cases III and IV, where we can see that the transmission from "$j$" can indirectly affect their second-nearest neighbors ($n_2$) through an intermediary person.}\label{fig.siq}
\end{figure}

At each time step $t \to t+1$, our SIQ model evolves with synchronous
updates as follows:
\begin{itemize}
\item Sub-step 1 (Transmission): All infected individuals at time $t$ transmit the disease to their susceptible neighbors with probability $\beta$. To simplify our theoretical equations, we will focus exclusively on the case in which the transmission probability $\beta$ is set to 1.
\item Sub-step 2 (Quarantine): Individuals who contracted the disease in the previous sub-step but also have regular access to testing, are immediately placed in quarantine along with their neighbors. 
\end{itemize}
In Fig.~\ref{fig.siq}, we present a few examples to illustrate how the states of the nodes change according to the rules of our model. This process of transmission (sub-step 1) and quarantine (sub-step 2) repeats iteratively until the system reaches a final stage in which individuals no longer change their state.

As indicated in the Introduction, is important to remark that for the specific case of $\beta=1$, our SIQ model is a simplified version of the SIRQ model studied in Ref.~\cite{valdez2023epidemic}. Therefore, several results reported in that work, such as the emergence of an abrupt phase transition, are also found in our SIQ model. However, the main results presented in Ref.~\cite{valdez2023epidemic} were primarily obtained from stochastic simulations on finite networks. Here, instead, thanks to our theoretical equations for $\beta=1$, we will be able to: 1) validate our simulation results, and 2) explore in more detail how our model behaves near the critical point, where stochastic fluctuations become significantly larger. On the other hand, while our primary focus is on the case in which $\beta=1$, it is worth noting that we have also explored scenarios in which $\beta<1$ through simulations. Our results (shown in Appendix~\ref{app.low1}) indicate that for high values of $\beta$ (though still less than 1), the system behaves qualitatively similar to the extreme case of $\beta=1$, displaying an abrupt phase transition. However, for low values of $\beta$, the system's behavior aligns more closely with a continuous transition.

In the following section, we present the core of our theoretical work for $\beta=1$, which is valid for factor graphs with a tree-like structure. This theoretical framework continues in Appendix~\ref{app.supp}. Then, in Sec.~\ref{sec.result} we provide a comparison between the theoretical results and our stochastic simulations on finite networks. Readers interested in a qualitative understanding of the results and their discussion can proceed directly to Sec.~\ref{sec.result}, returning to Sec.~\ref{sec.class} later for the detailed theoretical model.

\section{Mathematical formulation}~\label{sec.class}
Here we provide an overview of the generating functions, and time evolution equations used in our SIQ model, leaving the full technical details available in Appendix~\ref{app.supp}.

To develop our time evolution equations, we will use an effective-degree approach that has been applied in other epidemic models~\cite{miller2014epidemic, cai2014effective,lindquist2011effective}. In the context of an SIR model, this approach not only tracks the disease state of each node —susceptible, infected, or recovered— but also incorporates the states of their immediate neighbors~\cite{miller2014epidemic, cai2014effective,lindquist2011effective}. In our work, we will adapt this idea by considering cliques rather than individual nodes.  

Suppose that we randomly choose a clique at time $t$. We define two types of cliques:
\begin{itemize} 
\item Open cliques: are those cliques that fulfill both of the following conditions: i) no disease transmission has occurred yet among members, and ii) members with access to testing services still remain disease-free.
\item Closed cliques: are those in which one of the following events has occurred: i) at least one member has transmitted the disease within the clique (sub-step 1 of our model), or ii) at least one member with access to testing services has contracted the disease, which immediately triggers a quarantine order for the entire clique (sub-step 2). Once closed, a clique remains in this state throughout the epidemic process.
\end{itemize}
Fig.~\ref{fig.siq} illustrates how open cliques at time $t$ can transition to closed cliques at time $t+1$. In our work, we will show that by tracking  the time evolution of the density of open cliques and their composition, other magnitudes of our SIQ model can be calculated. To this end, we will first present the generating functions associated with open cliques and susceptible individuals (with and without access to testing). After that, we will obtain the basic reproduction number, the time evolution equations for the density of open cliques, and finally, the time evolution of other relevant quantities. Additionally, we make available in our GitHub repository~\cite{gith01}, the equations written in the Mathematica programming language.

\subsection{Generating functions for open cliques and susceptible individuals}\label{sec.class1}
\begin{figure}[ht]
\begin{center}
\begin{overpic}[scale=0.9]{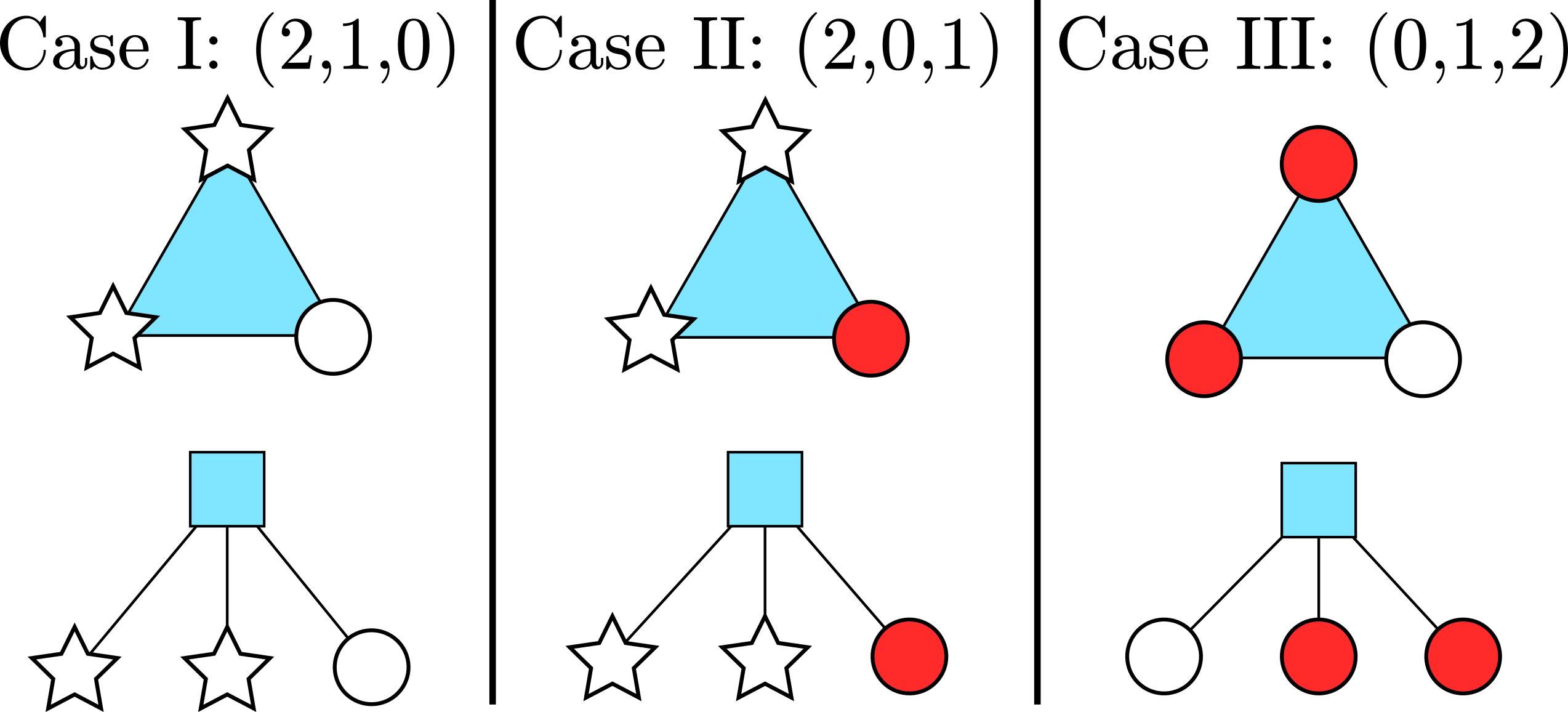}
\end{overpic}
\vspace{0cm}
\end{center}
\caption{Schematic figure displaying three examples of cliques with different compositions. Each clique contains ($\ell$,$r$,$i$) members, where $\ell$ denotes the number of people with regular access to testing, $r$ denotes the number of susceptible individuals without access to testing, and $i$ denotes the number of infected people without access to testing. For each clique, we show its bipartite representation. The symbols and colors used are consistent with those presented in Figs.~\ref{fig.cliq} and~\ref{fig.siq}.}\label{fig.plis}
\end{figure}

To describe the time evolution of the density of open cliques, we define $P^{t}_{\ell, r,i}$ as the fraction of open cliques at time $t$ containing the following:
\begin{itemize}
\item  $\ell$  susceptible members with access to testing,
\item  $r$ susceptible individuals without access to testing,
\item  $i$  infected individuals without access to testing.
\end{itemize}
In Fig.~\ref{fig.plis}, we show several examples of open cliques with various member configurations. For brevity, $P^{t}_{\ell,r,i}$ will be referred to as the fraction of open cliques containing $(\ell,r,i)$ members. Of course that, in the definition of $P^{t}_{\ell,r,i}$, we could also add another index to account for the number of quarantined people within an open clique (see for example Fig.~\ref{fig.siq} case III and IV where  $c_3$, which is an open clique, has a quarantined member at time $t+1$). However, we omit this index because quarantined individuals do not interact with other people, and therefore they are not relevant to the dynamics of disease spread. 

Based on this multivariate probability distribution $P_{\ell,r,i}^t$, we proceed to define the following generating functions which are just generalizations of the pgf's $F_0(x)$ and $F_1(x)$ given in Sec.~\ref{sec.netmod}:
\begin{itemize}     
\item $F_{0}^{t}(x,y,z)=\sum_{\ell}\sum_r\sum_i P_{\ell,r,i}^t x^{\ell} y^r z^i$. This function generalizes $F_{0}(x)$ (defined in Sec.~\ref{sec.netmod}), and corresponds to the generating function for the probability of randomly selecting an open clique with $(\ell,r,i)$ members. Note that $F_{0}^{t}(1, 1, 1)=\sum_{\ell}\sum_r\sum_i P_{\ell,r,i}^t \leq 1$ represents the total fraction of open cliques at time $t$.
\item $F_{1, \ell}^{t}(x,y,z)=\sum_{\ell}\sum_r\sum_i \frac{\ell P_{\ell,r,i}^t}{\kappa_{\ell}^t}x^{\ell-1} y^r z^i$. This function is analogous to $F_{1}(x)$ from Eq.~(\ref{eq.g01c}) and represents the generating function for the probability of selecting an open clique via a member who has regular access to testing services. Here, $\kappa_{\ell}^t=\sum_{\ell}\sum_r\sum_i \ell P_{\ell,r,i}^t$.
\item $F_{1, r}^t(x,y,z)=\sum_{\ell}\sum_r\sum_i \frac{r P_{\ell,r,i}^t}{\kappa_r^t}x^{\ell} y^{r-1} z^i$. This function is analogous to the previous one and represents the generating function for selecting an open clique via a susceptible member without access to testing. Here, $\kappa_r^t=\sum_{\ell}\sum_r\sum_i r P_{\ell,r,i}^t$.
\end{itemize}

Moving forward, we now introduce the generating functions for the subpopulation of susceptible individuals who have regular access to testing. These functions, denoted $G_{0,\ell}^t(x)$ and $G_{1,\ell}^t(x)$, are analogous to the functions $G_0(x)$ and $G_1(x)$ given in Sec.~\ref{sec.netmod}: 
\begin{eqnarray}
G_{0,\ell}^t(x) &=& \sum_{k_I=0}^{\infty} S_{\ell}(k_I,t) x^{k_I},\\
G_{1,\ell}^t(x) &=& \sum_{k_I=1}^{\infty}  \frac{k_I S_{\ell}(k_I,t)}{\langle k_{\ell}\rangle} x^{k_I-1},
\end{eqnarray}
where $\langle k_{\ell}\rangle=\sum_{k_I=0}^{\infty} k_I S_{\ell}(k_I,t)$, and $S_{\ell}(k_I,t)$ represents the proportion of susceptible individuals at time $t$ who have  regular access to testing and with a membership $k_I$ (meaning that an individual has $k_I$ cliques in the factor graph). Note that evaluating $G_{0,\ell}^t(x)$ at $x=1$, provides the total fraction of susceptible people with access to testing services at time $t$, i.e., $G_{0,\ell}^t(1)=S_{\ell}^t$.

Similarly, for the subpopulation of susceptible individuals without access to testing, we define the analogous generating functions:
\begin{eqnarray}
G_{0,r}^t(x) &=& \sum_{k_I=0}^{\infty}  S_{r}(k_I,t) x^{k_I},\\
G_{1,r}^t(x) &=& \sum_{k_I=1}^{\infty}  \frac{k_I S_{r}(k_I,t)}{\langle k_{r}\rangle} x^{k_I-1},
\end{eqnarray}
where $\langle k_{r}\rangle=\sum_{k_I=0}^{\infty} k_I S_{r}(k_I,t)$, and evaluating $G_{0,r}^t(x)$ at $x=1$ provides the total fraction of the population without access to testing at time $t$ (i.e., $G_{0,r}^t(1)=S_{r}^t$).

Now, as in Sec.~\ref{sec.netmod}, we can employ the power property of generating functions to obtain the pgf's describing the local neighborhood of an individual. As an example,  consider that, at time $t$, we randomly select a susceptible individual "$j$" who doesn't have access to testing. The neighborhood composition of this person is captured by
\begin{eqnarray}\label{eq.g1neighreg}
    \mathcal{G}^{(1)}_{r}(x_1,y_1,z_1) &=& G_{0,r}^t(F_{1 , r}^{t}(x_1,y_1,z_1))=\sum_{\ell^{\dag}=0}\sum_{r^{\dag}=0}\sum_{i^{\dag}=0}a_{\ell^{\dag}r^{\dag}i^{\dag}}x_1^{\ell^{\dag}}y_1^{r^{\dag}}z_1^{i^{\dag}},
\end{eqnarray}
where $a_{\ell^{\dag}r^{\dag}i^{\dag}}$ is the probability that the chosen person has in their neighborhood: 1) $\ell^{\dag}$ susceptible individuals with access to testing services, 2) $r^{\dag}$ susceptible people without access to these services, and 3) $i^{\dag}$ infected people without access to these services. This pgf generalizes the probability-generating function presented in Eq.~(\ref{eq.gvec1}). An interesting property of this generating function, which we will use extensively in our time evolution equations, is that by setting $z_1=0$, we are only considering those configurations where "$j$" has no infected first-nearest neighbors. Conversely, by setting $z_1=1$, we count all configurations in which "$j$" has any number of infected first-nearest neighbors. Similar interpretations can be made for $x_1=y_1=0$ and $x_1=y_1=1$.

On the other hand, if we randomly choose "$j$" through a clique, the corresponding pgf for the excess-degree distribution (discriminated by type) is given by:
\begin{eqnarray}\label{eq.g1neighsusreg1}
   \mathcal{G}^{(1)}_{1, r}(x_1,y_1,z_1) &=& G_{1,r}^t(F_{1 , r}^{t}(x_1,y_1,z_1)),
\end{eqnarray}
which is a generalization of Eq.~(\ref{eq.gvec1prim}).

Analogously, the probability-generating functions that describe the neighborhood of individuals who have regular access to testing are:
\begin{eqnarray}\label{eq.g1neighlead}
    \mathcal{G}^{(1)}_{\ell}(x_1,y_1,z_1) &=& G_{0,\ell}^t(F_{1 , \ell}^{t}(x_1,y_1,z_1)),\\
    \mathcal{G}^{(1)}_{1, \ell}(x_1,y_1,z_1) &=& G_{1,\ell}^t(F_{1 , \ell}^{t}(x_1,y_1,z_1)).
\end{eqnarray}
\subsection{Basic reproduction number}\label{sec.RepNum}

Using the generating functions defined in the previous sections, we can now estimate the basic reproduction number $R_0$ at time $t=0$ for a scenario in which the initial fraction of the infected population is microscopic. Because our SIQ model for $\beta=1$ is a special case of the SIRQ model explored in \cite{valdez2023epidemic}, the derivation of $R_0$ presented in that work remains valid here. Following Refs.~\cite{valdez2023epidemic,valdez2024explosive}, $R_0$ is calculated as the ratio between the number of infected second-nearest neighbors, and the number of infected first-nearest neighbors. The expression of $R_0$ is given by:
\begin{eqnarray}\label{eq.r00}
    R_0=\left(\frac{dF_1((1-f)x)}{dx}\Bigr|_{\substack{x=1}} \frac{d\mathcal{G}^{(1)}_{1}(x)}{dx}\Bigr|_{\substack{x=1}}\right)/\left( \frac{dF_1(x)}{dx}\Bigr|_{\substack{x=1}}\right).
\end{eqnarray}
To understand this expression, consider that we randomly choose an individual (referred to as the index-case) through a clique and assume that this person is initially infected. The denominator in Eq.~(\ref{eq.r00}) represents the average number of susceptible first-nearest neighbors who can be infected by this index-case. Once infected, those individuals without access to testing (represented by the term $\frac{dF_1((1-f)x)}{dx}\Bigr|_{\substack{x=1}}$) will transmit the disease further to their own neighbors. Each of these untested individuals will, on average, transmit the disease to $\frac{d\mathcal{G}^{(1)}_{1}(x)}{dx}\Big|_{x=1}$ additional people. So the numerator represents the total average number of second-nearest neighbors of the index-case who will contract the disease.

\subsection{Time-discrete equations for open cliques}\label{sec.timediscMar} 
Here, we will formulate the time-evolution equations for $P_{\ell,r,i}^t$, as well as those for $S_{r}(t,k_I)$ and $S_{\ell}(t,k_I)$. To this end, we will begin by listing the transitions that affect the number and composition of open cliques.

On one hand, it is important to note that any open clique containing at least one infected member at time $t$ will become a closed clique at $t+1$. Figure~\ref{fig.siq} illustrates this transition: for example, in case I, cliques $c_1$ and $c_2$ transition from open to closed because they contain an infected member who spreads the disease. Mathematically, open cliques containing $(\ell^*,r^*,i^*)$ members with $i^*\geq 1$ will exit the compartment of open cliques during the time step $t\to t+1$, and become closed cliques.

On the other hand, for open cliques containing $(\ell^*,r^*,i^*=0)$ members at time $t$, the following transitions preserve the open status of these cliques:
\begin{itemize}
    \item Among susceptible members who have access to testing, a portion $\ell \leq \ell^*$  remains susceptible, while the rest $\ell^*-\ell$ transition to a quarantine state (because they receive quarantine orders from other cliques). The probability of this event is denoted as $p(\ell|\ell^*)$. Figure~\ref{fig.siq}-case IV gives an example of this transition, where a member of clique $c_3$ who has regular access to testing, is placed in quarantine during the time step $t\to t+1$.
    \item Among susceptible members without access to testing: 1) a number $i$ contract the disease from other cliques (see for example, clique $c_1$ in Fig.~\ref{fig.siq}-Case I), 2) a number $r$ remain susceptible, and 3) a number $r^*-i-r$ are quarantined during the time step $t \to t+1$. The probability of this event is denoted as $p(i,r|r^*)$.
\end{itemize}
With these transition probabilities, we can now write the Markovian equations governing the fraction of open cliques containing $(\ell,r,i)$ members as,
\begin{eqnarray}\label{eq.markovv}
P_{\ell,r,i}^{t+1} = \sum_{\ell^*\geq \ell} \sum_{r^*\geq r+i}  P_{\ell^*,r^*,0}^t \times p(\ell|\ell^*)  p(i,r|r^*),
\end{eqnarray}
where $p(\ell|\ell^*)  p(i,r|r^*) \equiv p(\ell,r,i|\ell^*,r^*,0)$ can be considered as a transition probability tensor. The exact analytical expressions for $p(\ell|\ell^*)$  and $p(i,r|r^*)$ are given by,
\begin{eqnarray}
    p(\ell|\ell^*)&=& \binom{\ell^*}{\ell} \Phi^{\ell} \times (\mathcal{G}^{(1)}_{\ell,1}(1,0,1)-\Phi)^{\ell^*-\ell},\label{eq.pellell}\\
    p(i,r|r^*)&=&\binom{r^*}{i,r,r^*-i-r}\sigma^i \Psi^r \times (1-\sigma-\Psi)^{r^*-i-r},\label{eq.pirell}
\end{eqnarray}
with,
\begin{eqnarray}
    \Phi&=&\mathcal{G}^{(1)}_{1, \ell}[\mathcal{G}^{(1)}_{1, \ell}(1,1,0),1,0],\\
    \Psi&=&\mathcal{G}^{(1)}_{1, r}[\mathcal{G}^{(1)}_{1, \ell}(1,1,0),1,0],\\
    \sigma &=& G_{1,r}^t[F_{1, r}^t(0,1,1) - F_{1, r}^t(0,1,0) + F_{1, r}^t(\mathcal{G}^{(1)}_{1, \ell}(1,1,0),1,0) ] - \Psi,
\end{eqnarray}
and their derivation is explained in detail in Appendix~\ref{app.supp}. An important observation is that unlike random networks without cliques, where the dynamics can be described by a single time-evolution equation~\cite{miller2011note}, in our work, instead, the dynamics is governed by the system of equations given in~(\ref{eq.markovv}) which contains an order of $\mathcal{O}((k_{C,max})^3)$ equations, where $k_{C,max}$ is the size of the largest clique. While this clearly increases the complexity of the model, our system of equations will allow us to accurately describe the behavior of our SIQ model on networks with cliques in the thermodynamic limit.

Having computed $P_{\ell,r,i}^{t+1}$, we then proceed to calculate the new proportions of susceptible individuals with membership $k_I$. Specifically, for susceptible individuals without access to testing services, the value of $S_{r}(t+1, 
k_I)$ is determined as follows:
\begin{eqnarray}
    S_{r}(t+1, k_I) &=& S_{r}(t, k_I)\left(F_{1, r}^t\left[\mathcal{G}^{(1)}_{1, \ell}(1,1,0),1,0 \right]\right)^{k_I},
\end{eqnarray}
where $F_{1, r}^t\left[\mathcal{G}^{(1)}_{1, \ell}(1,1,0),1,0 \right]$ is the probability that a clique remains open at time $t+1$, as explained in more detail in Appendix~\ref{app.supp}. Similarly, for a susceptible individual with access to testing services, $S_{\ell}(t+1, k_I)$ is calculated as,
\begin{eqnarray}
 S_{\ell}(t+1, k_I) &=&  S_{\ell}(t, k_I)\left(F_{1, \ell}^t\left[\mathcal{G}^{(1)}_{1, \ell}(1,1,0),1,0 \right]\right)^{k_I}.
\end{eqnarray}

\subsection{Time-discrete equations for other epidemiological magnitudes}

Once $P_{\ell,i,s}^{t+1}$, $S_{\ell}(t+1, k_I)$ and $S_{r}(t+1, k_I) $ are calculated, we proceed to compute other epidemiological quantities that describe the disease spread in our SIQ model. Particularly, we will focus on: 1) the total fraction of susceptible people, denoted as $S(t+1)$ and  2) the total fraction of people who contracted the disease during the time step $t\to t+1$, denoted as $I_{new}(t+1)$.

On one hand, $S(t+1)$ is obtained by simply summing all susceptible individuals with and without access to testing:
\begin{eqnarray}
S(t+1)&=&\sum_{k_I} \left(S_{\ell}(t+1, k_I)+S_{r}(t+1, k_I)\right).
\end{eqnarray}

On the other hand, $I_{new}(t+1)$ is calculated using the following equation:
\begin{eqnarray}
    I_{new}(t+1)&=& \mathcal{G}^{(1)}_{\ell}\left[1,1,1 \right]- \mathcal{G}^{(1)}_{\ell}\left[1,1,0 \right]+\nonumber\\
    &&+\mathcal{G}^{(1)}_{r}\left[1,1,1 \right]- \mathcal{G}^{(1)}_{r}\left[1,1,0 \right],
\end{eqnarray}
where 
\begin{itemize}
    \item $ \mathcal{G}^{(1)}_{\ell}\left[1,1,1 \right]- \mathcal{G}^{(1)}_{\ell}\left[1,1,0 \right]$ represents the fraction of regularly tested susceptible individuals at time 
$t$ with at least one infected first-neighbor,
    \item $\mathcal{G}^{(1)}_{r}\left[1,1,1 \right]- \mathcal{G}^{(1)}_{r}\left[1,1,0 \right]$ represents the fraction of susceptible people without access to testing services and with at least one infected first-neighbor.
\end{itemize}

After computing $S(t+1)$ and $I_{new}(t+1)$, we proceed to update all the generating functions defined in the previous section that depend on $P_{\ell,i,s}^{t}$, $S_{\ell}(t, k_I)$ and $S_{r}(t, k_I)$, and finally, time is increased by 1.

\section{Results}\label{sec.result}
\subsection{Final Stage}\label{sec.resultfinal}
In this section, we will numerically compare our theoretical predictions with the results of numerical simulations under the same initial conditions and parameter settings. In particular, we will focus here on networks with cliques where the degree distributions $P(k_C)$ and $P(k_I)$ follow a truncated Poisson function,
\[
Pois(\lambda,k_{min},k_{max}) = 
\begin{cases} 
      c \frac{\lambda^k\exp(-\lambda)}{k!}, & \text{if } k_{min}\leq k \leq k_{max} \\
      0, & \text{otherwise}
\end{cases}
\]\label{eq.trunPois}
where $c$ is a normalization constant. Specifically, we will use $P(k_C)\sim Pois(7,2,20)$ and $P(k_I)\sim Pois(3,1,20)$ (hereafter referred to as "ER$_7$-ER$_3$" networks for brevity). On the other hand, as initial conditions, we assume that a fraction  $\epsilon=10^{-6}$ of the population without access to testing is infected, while the rest of the population is susceptible.

\begin{figure}[H]
\begin{center}
\begin{overpic}[scale=0.60]{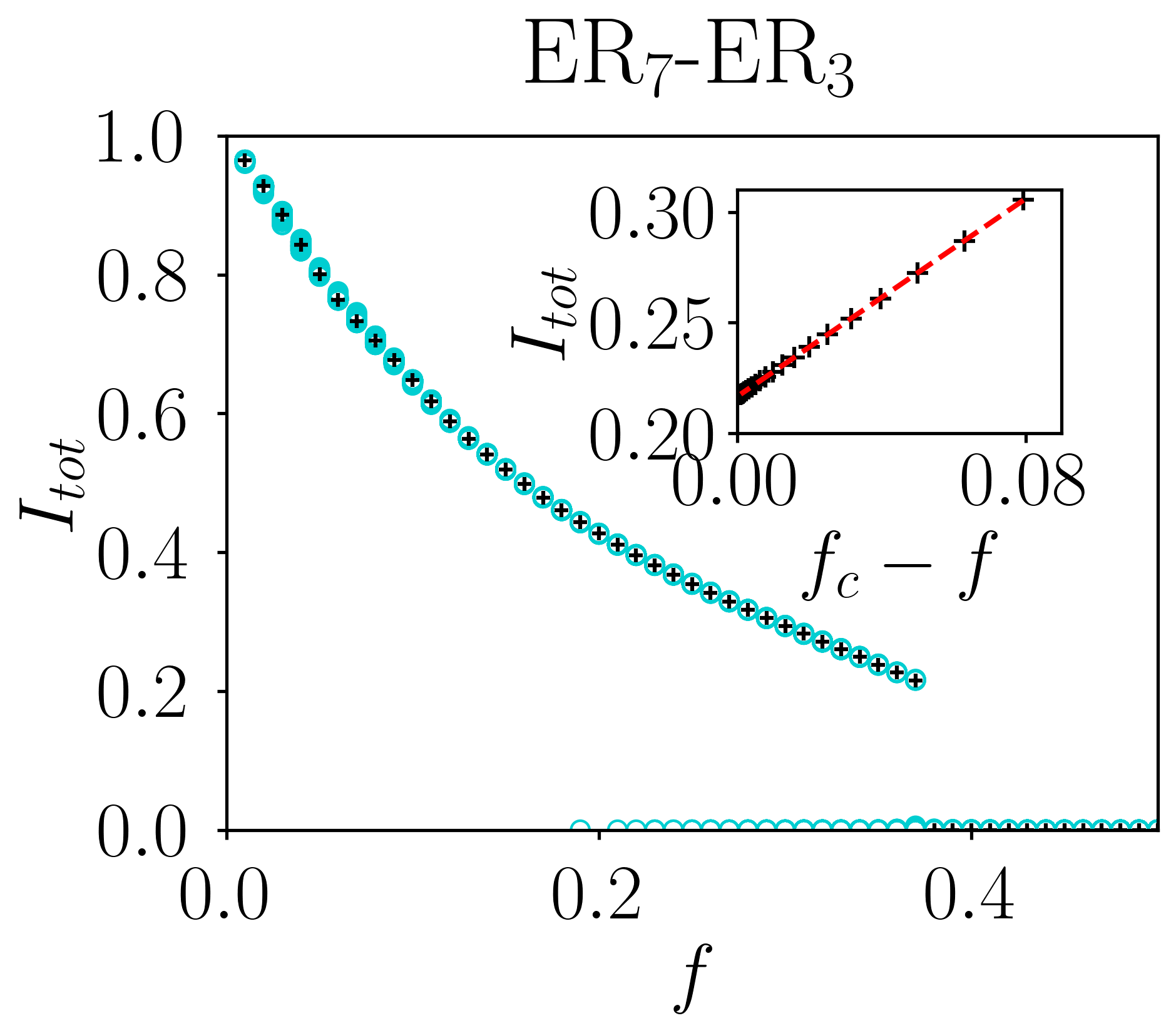}
  \put(85,18){}
\end{overpic}
\vspace{-1.1cm}
\end{center}
\caption{Fraction of people who have ever been infected $I_{tot}$ as a function of $f$ for a network with cliques where $P(k_C)\sim Pois(7,2,20)$ and $P(k_I)\sim Pois(3,1,20)$. The value $f_c=0.369$, obtained when $R_0=1$ in Eq.~(\ref{eq.r00}), marks the epidemic threshold. In the main plot, the symbols "$+$" correspond to the final fraction $I_{tot}$ obtained by integrating our theoretical equations presented in Sec.~\ref{sec.class}, while the circles correspond to a scatter plot obtained from stochastic simulations for 100 realizations on networks with $N_I=10^7$. The inset displays $I_{tot}$ vs $f_c-f$  obtained from our theoretical equations (indicated by "$+$"), with a dashed line representing a linear fit.}\label{fig.steady}
\end{figure}

In Fig.~\ref{fig.steady}, we display $I_{tot}$ at the final stage as a function of $f$. Here, $I_{tot}$ represents the fraction of the population that have ever been infected and mathematically is computed by summing the number of new infections at each time step: $I_{tot}=\sum_{t=0}^{\infty} I_{new}(t)$. From this figure, we can see that the agreement between theory and simulations is excellent. In Appendix~\ref{app.bet1}, we further demonstrate how the stochastic simulations converge to the theoretical solution as the network size increases. As intuitively expected, both theory and simulations show that as more people have access to testing (higher $f$), the overall fraction of the infected population ($I_{tot}$) decreases, or in other words, it becomes easier to contain the spread of the disease.

In addition, from the figure we observe that $I_{tot}$ is rapidly suppressed around $f_c\approx 0.369$. This sharp behavior is consistent with earlier observations made in our previous study of the SIRQ model~\cite{valdez2023epidemic}, where we noticed from stochastic simulations an abrupt transition around $f=f_c$. However, it is important to note that in the SIRQ model, we did not investigate the behavior of epidemics near this threshold due to the high computational cost of stochastic simulations.

Now that we have the time evolution equations for our SIQ model, we can more precisely investigate this transition by numerically integrating our equations and observing how the system behaves as $f$ approaches $f_c$. In contrast to hybrid transitions which require a supercritical behavior with diverging slope of the order parameter~\cite{d2019explosive}, from the inset of Fig.~\ref{fig.steady}, we can see that in the limit $f\to f_c$, $I_{tot}$ behaves like a linear function $I_{tot} = a_i - b_i(f - f_c)$ where $a_i$ and $b_i$ are fitting parameters. Consequently, the absence of 
diverging slope in $I_{tot}$ indicates that our SIQ model does not undergo a hybrid transition, but rather a "Type II" abrupt phase transition (see Ref.~\cite{d2019explosive}). We also provide additional results in Appendix~\ref{app.bet1} to show that this observation holds true for a variety of network configurations as well.

\subsection{Time evolution}
Having established the agreement between our theoretical framework and stochastic simulations for the final stage, we now turn our attention to the temporal dynamics of the SIQ model.

\begin{figure}[ht]
\begin{center}
\begin{overpic}[scale=0.45]{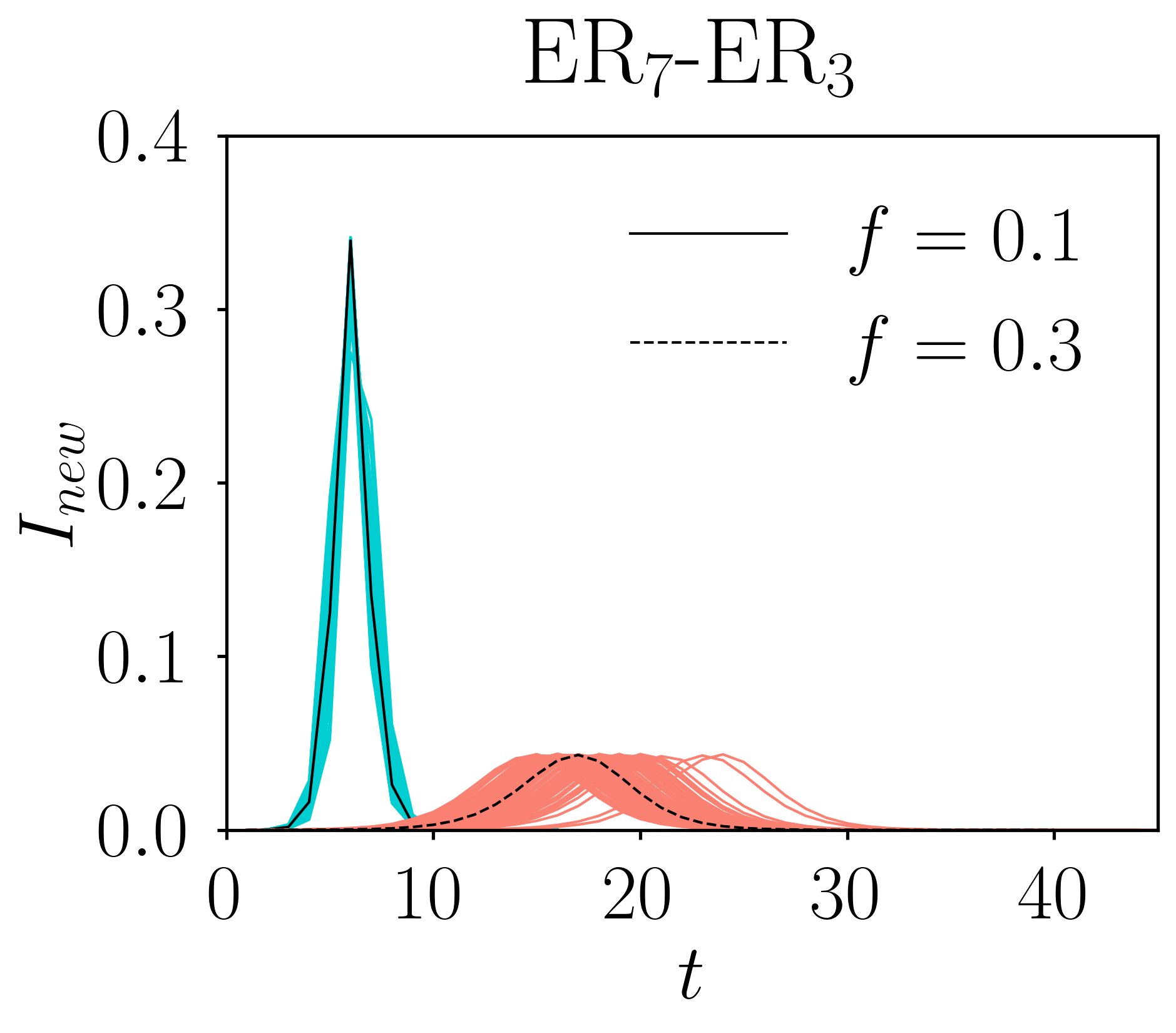}
  \put(85,38){(a)}
\end{overpic}
\vspace{-1.1cm}
\begin{overpic}[scale=0.45]{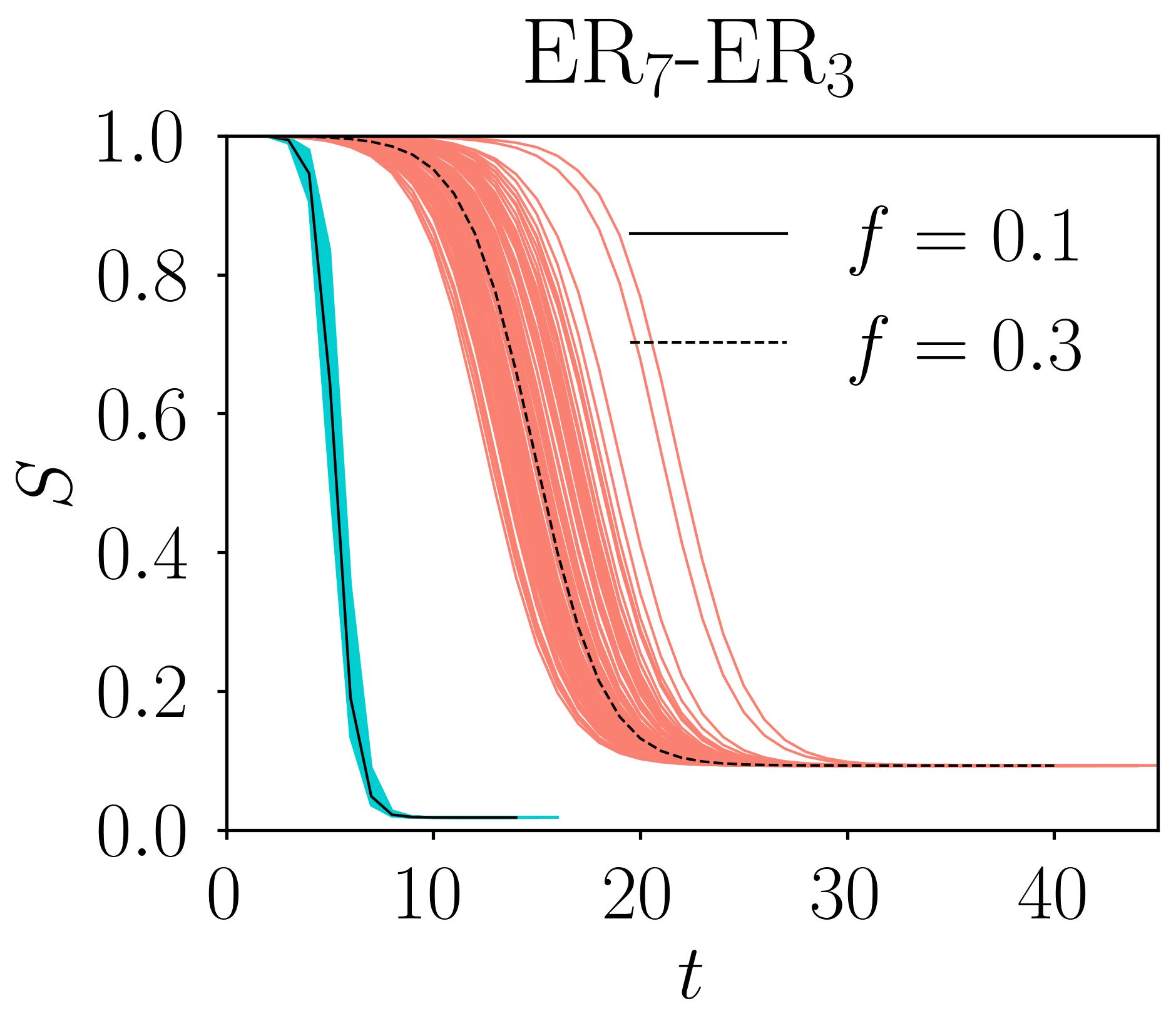}
  \put(85,38){(b)}
\end{overpic}
\end{center}
\caption{Time evolution of $I_{new}$ (panel a) and $S$ (panel b) for an ER$_7$-ER$_3$ network. Colored lines correspond to 100 stochastic realizations for $N_I=10^7$ and: $f=0.1$ (light blue), $f=0.3$ (salmon). On the other hand, black lines correspond to the theoretical solutions obtained from our equations presented in Sec.~\ref{sec.class}.} \label{fig.timeevol}
\end{figure}

Figures~\ref{fig.timeevol}a-b display the time evolution of $I_{new}$ and $S$ for $f=0.1$ and $f=0.3$ on ER$_7$-ER$_3$ networks. From these figures, one can observe that the theoretical predictions are in excellent agreement with our numerical simulation results. Additionally, in Appendix~\ref{app.bet1}, we show that our equations also accurately predict the time evolution of the SIQ model for other network topologies.

\begin{figure}[H]
\begin{center}
\begin{overpic}[scale=0.45]{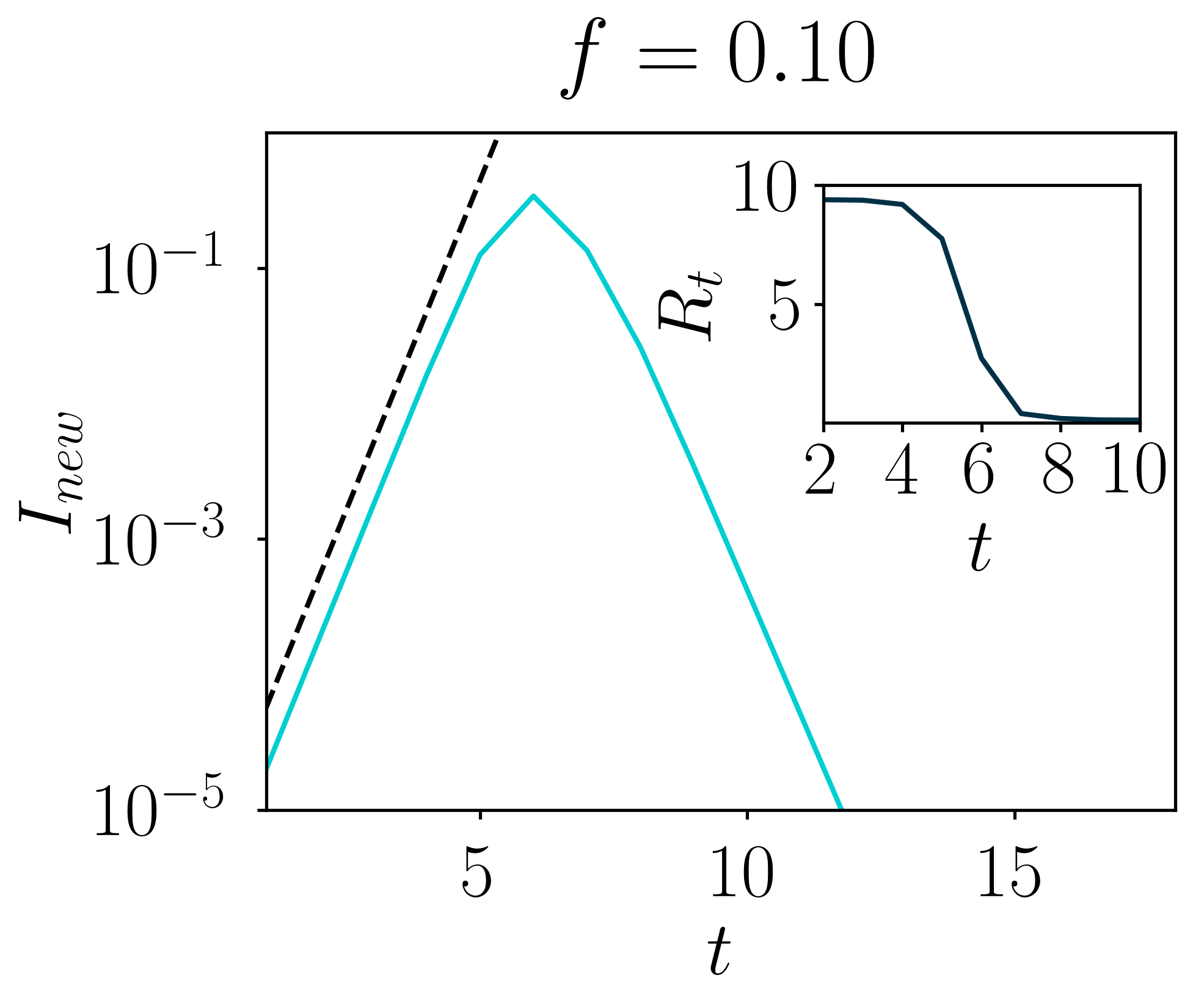}
  \put(85,28){(a)}
\end{overpic}
\begin{overpic}[scale=0.45]{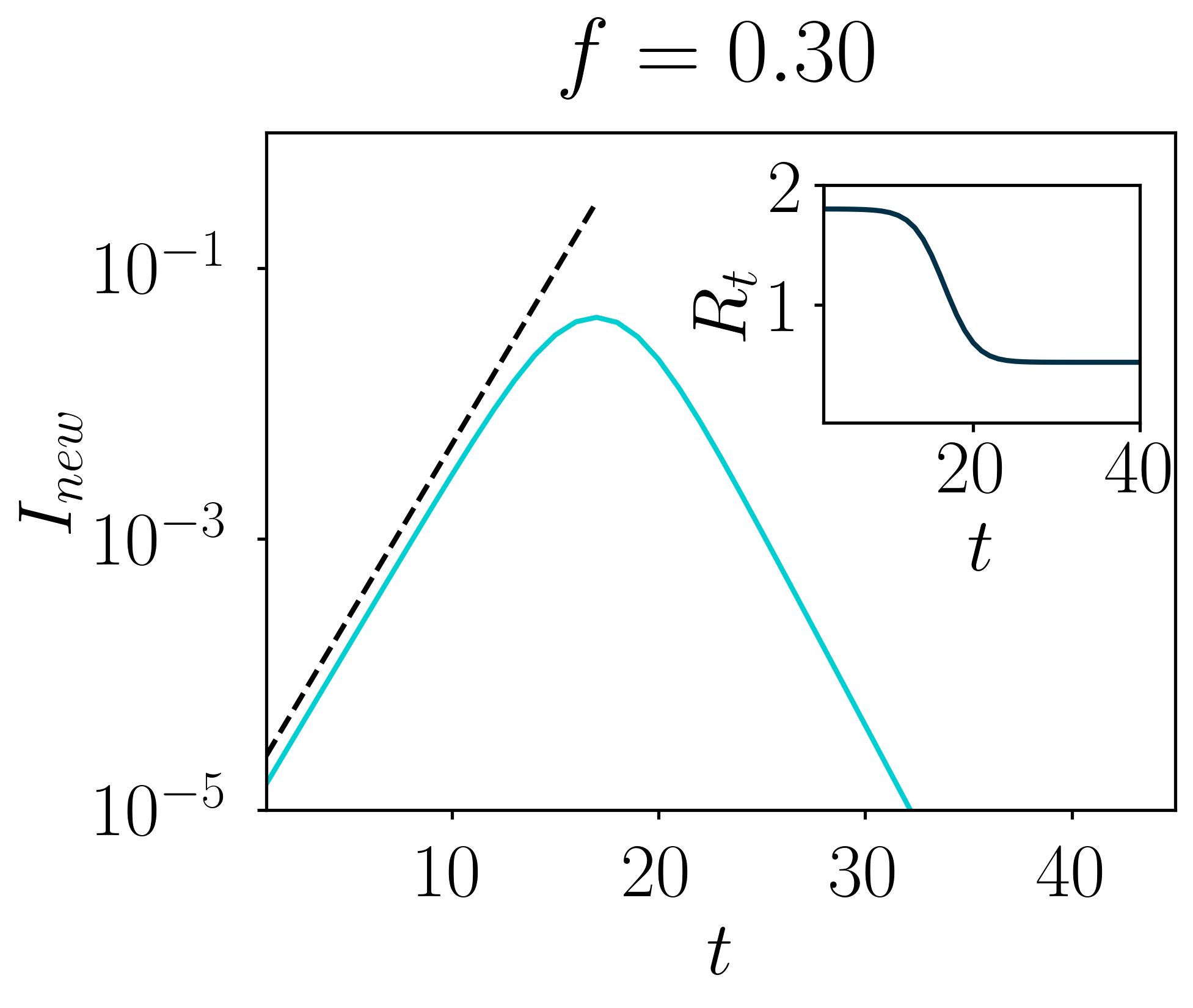}
  \put(85,28){(b)}
\end{overpic}
\begin{overpic}[scale=0.45]{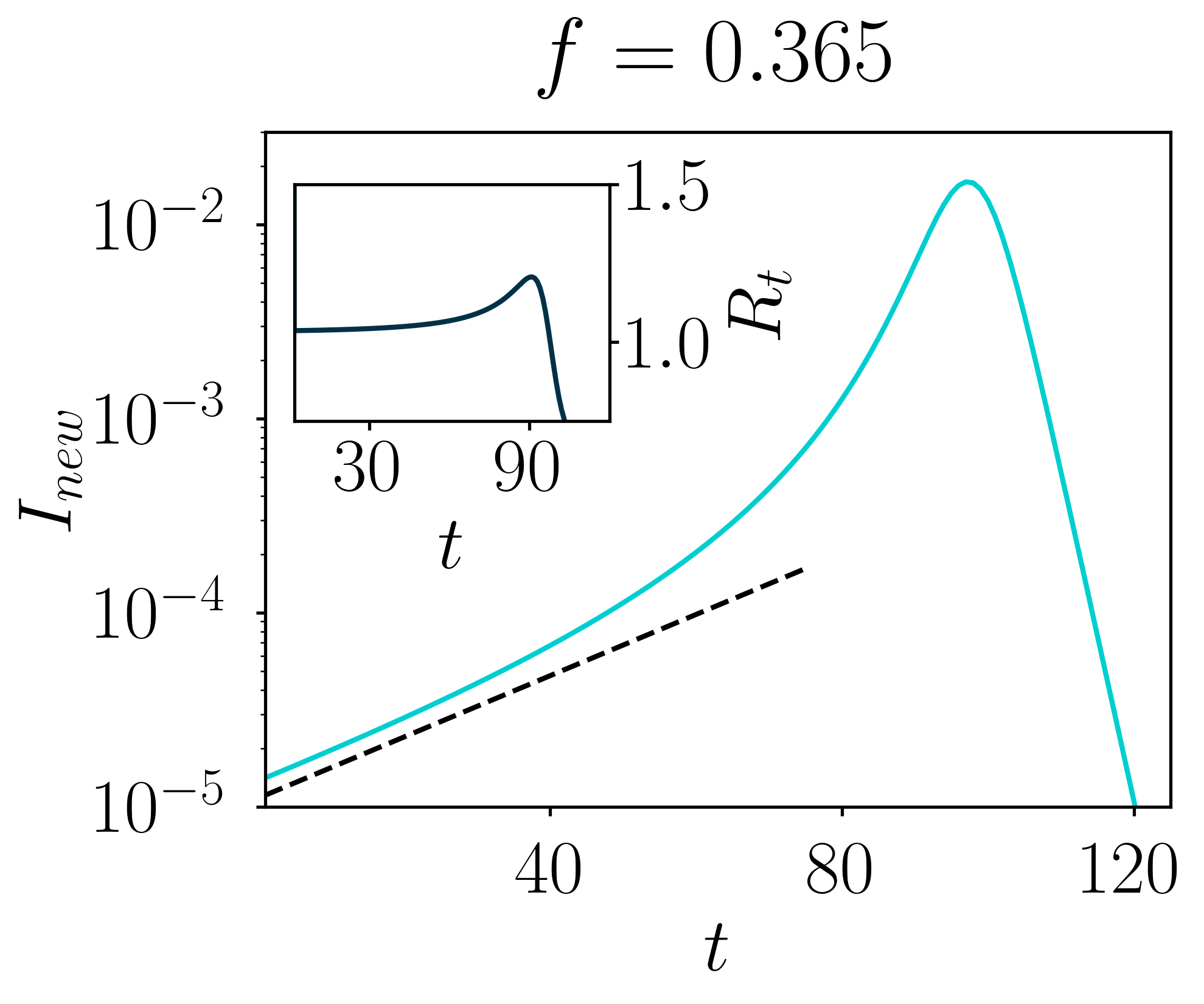}
  \put(85,58){(c)}
\end{overpic}
\begin{overpic}[scale=0.37]{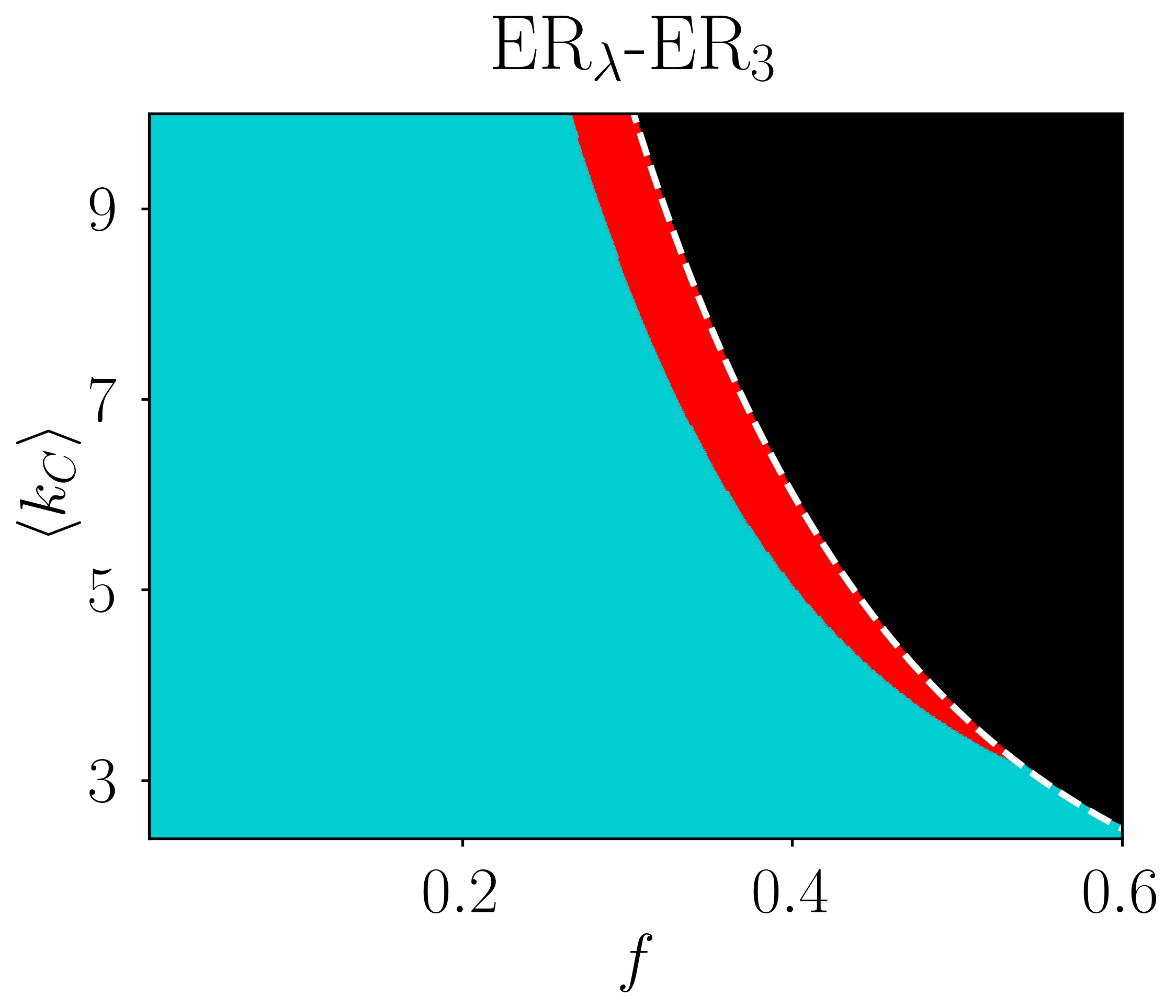}
  \put(25,58){(d)}
\end{overpic}
\vspace{-1.1cm}
\end{center}
\caption{Panels a-c: Time evolution of $I_{new}$ for ER$_7$-ER$_3$ networks and for different values of $f$. The vertical axis is plotted on a logarithmic scale, while the horizontal axis is linear. Solid lines correspond to our theoretical solutions obtained from the equations in Sec.~\ref{sec.class}. Each dashed line is an exponential function $\propto \exp(\alpha t)$ where $\alpha=\ln(R_0)$ and $R_0$ is the basic reproduction number given by Eq.~(\ref{eq.r00}). The insets show the time evolution of the effective reproduction number $R_t$ on a linear scale for both axes. Panel d: Phase diagram in the $\langle k_C\rangle$-$f$ plane, where clique sizes follow a truncated Poisson distribution Pois($\lambda$,2,20) with $\lambda \in (1,11)$, and $\langle k_C\rangle$ represents the average clique size. The light blue region corresponds to the area in the plane where the super-exponential growth dynamics is not detected. The red region indicates super-exponential growth, while the black region is free of epidemics. The white dashed line represents the threshold where the basic reproduction number equals 1 (see Eq.(\ref{eq.r00})).}\label{fig.super}
\end{figure}

Moving forward, we will now explore the temporal evolution of infections across different values of $f$ in more detail. From Figs.~\ref{fig.super}a-c, we notice that $I_{new}(t)$ initially increases over time as an exponential function.  Moreover, the exponential growth rate $\alpha$ obeys the relationship $\alpha = \ln(R_0)$, where $R_0$ is the basic reproduction number defined in Eq.~(\ref{eq.r00}). This relationship can be understood by examining the early stages of the epidemic, when the vast majority of the population is still susceptible.  During this period, the transmission process can be approximated by a Galton-Watson branching process~\cite{spouge2019accurate} where each infected individual transmits the disease, on average, to $R_0$ members of the population. Because the infected population is still small at this point, we can safely ignore the probability of a susceptible person receiving the infection from multiple sources simultaneously. As a result, during this period, the number of new infections grows as $(R_0)^t$, which can be expressed equivalently as an exponential function: $(R_0)^t=\exp(\alpha t)$ with $\alpha = \ln(R_0)$.

After this initial regime where $I_{new}$ increases exponentially, we observe that the growth of $I_{new}(t)$ gradually slows down. As illustrated in Figs.~\ref{fig.super}a and b, for values of $f$ far from the critical threshold $f_c \cong 0.369$, $I_{new}(t)$ follows the typical trajectory observed in many epidemic models: there is an initial exponential growth, and then a well-defined peak and a subsequent decline. This behavior is also confirmed by the effective reproduction number $R_t$ (estimated as $R_t = I_{\text{new}}(t) / I_{\text{new}}(t-1)$). As shown in the insets of Figs.~\ref{fig.super}a-b, $R_t$ decreases steadily, which demonstrates that for $f$ values far from $f_c$, the propagation starts to slow down from the very beginning of the epidemic.

However, as $f$ approaches its critical value $f_c$, the epidemic dynamics deviates from this typical growth pattern. As illustrated in Fig.~\ref{fig.super}c, following an initial exponential increase (with a rate $\alpha = \ln(R_0)$), we can see that the fraction of new infections accelerates even further.  This accelerated growth is also reflected in the time evolution of $R_t$, as shown in the inset of Fig.~\ref{fig.super}c where we can observe that $R_t$ increases during the early stages of the outbreak. 

One possible explanation for this phenomenon is that our SIQ model becomes less effective over time because healthy individuals with regular access to testing are gradually removed from the network. A concrete example of this process is shown in Fig.~\ref{fig.siq}, case IV, where we can see that during the time step $t\to t+1$, clique $c_3$ loses its only member who can monitor and respond to infections. As a result of this event, subsequent infections within $c_3$ will go undetected, allowing the disease to spread more freely through the network. In other words, this example shows that our quarantine strategy has a self-undermining effect that gradually allows the disease to spread more easily. However, it is important to note that this mechanism alone is not enough to explain the super-exponential growth because, as we will see in Fig.~\ref{fig.super}d, networks without cliques do not exhibit this phenomenon.

To investigate the role of cliques in the emergence of this super-exponential growth, we will now explore the time evolution of the effective reproduction number $R_t$ in the $\langle k_c \rangle$-$f$ parameter space, where $\langle k_c\rangle$ is the average clique size of an "ER$_{\lambda}$-ER$_3$" network. Specifically, for each point in this parameter space (where $R_0>1$), we numerically integrate our time evolution equations for 10 time steps and analyze the behavior of $R_t$ from $t=3$ to $t=10$~\footnote{In our implementation, we integrate the equations from time $t = 3$ to $t = 10$ while ensuring that $I_{\text{new}}$ continues to increase. If $I_{\text{new}}$ ceases to grow during this period, we halt the integration.}. If $R_t$ monotonically decreases during this period, we will say that $I_{new}(t)$ exhibits the standard exponential growth. On the other hand, if $I_{new}(t)$ does not decrease monotonically, this is an indication that the epidemic growth is faster than exponential. Our results shown in Fig.~\ref{fig.super}d reveal that this accelerated behavior becomes more apparent near the critical curve where $R_0=1$, and consequently, this suggests that the emergence of accelerated transmission dynamics could serve as an indication that the system is close to the transition point. On the other hand, from the figure we observe that the phenomenon of faster-than-exponential growth seems to be absent in networks with smaller clique sizes ($\langle k_c\rangle$). Therefore, our results indicate that larger cliques play a crucial role in facilitating the acceleration of disease spread. In Appendix~\ref{app.bet1}, we show that these observations qualitatively hold for other networks with cliques with different distributions.

Beyond the temporal dynamics, Fig.~\ref{fig.super}(d) also reveals that as $\langle k_c\rangle$ increases, the value of $f_c$ decreases, or in other words, fewer people need to be tested to prevent an epidemic (see also Figs.~\ref{fig.suppHeat}a-b in Appendix~\ref{app.bet1}). Qualitatively similar results were obtained by Rizi et al.~\cite{rizi2024effectiveness}, whose work showed that contact tracing becomes more effective at preventing epidemics in networks with larger cliques. This is particularly significant, especially considering that larger groups, in the absence of any intervention, are known to increase the probability of epidemics~\cite{caillaud2013epidemiological,danon2021household}. Therefore, our results, along with those of Rizi et al.~\cite{rizi2024effectiveness}, indicate that some non-pharmaceutical interventions could reverse this vulnerability, turning the structural property that normally increases epidemic risk into an advantage for containment strategies.

\section{Conclusions}\label{sec.conclu}

To summarize, in this work, we have studied the dynamics and the final stage of an SIQ model on networks with cliques. In particular, for this model, we derived the dynamic equations governing the time evolution of disease propagation (for $\beta=1$), which allowed us to explore in detail how an epidemic spreads on networks with complete sub-graphs in the thermodynamic limit.

On one hand, from our theoretical equations, we found that cliques can induce an abrupt epidemic transition around a critical value $f_c$. This result is consistent with our previous findings reported in Ref.~\cite{valdez2023epidemic}, where such abrupt transitions were suggested based on numerical simulations on finite networks. Additionally, from our theoretical equations, we obtained that the final fraction of infected individuals decreases linearly with $f$ (in the limit of $f\to f_c$), and therefore our SIQ model undergoes an abrupt transition similar to "Type II" transitions described in~\cite{d2019explosive}. 

On the other hand, our results on the time evolution of the SIQ model revealed that, as $f$ approaches $f_c$, the number of new cases grows faster than an exponential function, suggesting that the emergence of this accelerated transmission dynamics could serve as an indication that the system is close to the transition point. For the networks explored in this investigation, our results indicate that this accelerated behavior tends to occur in networks with larger cliques.

While the theoretical framework developed in this work has allowed us to study the spread of epidemics in the thermodynamic limit, it should be noted that it is currently restricted to a discrete-time dynamics and $\beta=1$. Future research might explore whether a continuous-time version of the SIQ model exhibits similar abrupt transitions and super-exponential growth. For such an extension, we could consider both Markovian and non-Markovian transmission and detection processes. 

Additional research directions could explore quarantine strategies that isolate not only the detected individual and their first-neighbors but also their second- or even higher-order neighbors, to evaluate the impact of broader isolation measures on the epidemic dynamics. Another potential line of research could investigate alternative ways of distributing testing access across the network. In our current model, we assumed that each individual has the same probability $f$ of having access to testing. A natural extension would be to consider scenarios in which the probability of having access to testing depends on the membership $k_I$. Alternatively,  drawing inspiration from statistical physics models of hard-sphere gases on networks with cliques (see Refs.~\cite{weigt2003glassy,hansen2005hard,hansen2005harddyn}), we could explore a scenario in which people with access to testing are arranged periodically as in a crystalline structure or aperiodically as in a liquid state or a glassy state. We hope that the model and theoretical framework developed here will contribute to the study of other dynamic processes in complex networks with cliques.

\section{Acknowledgements}\label{Sec.Ack}
This work was partially funded by ANPCyT (PICT-2021-I-INVI-00255), UNMdP  (EXA 1193/24), and CONICET.

\appendix


\section{Supplementary mathematical details for $\beta=1$}\label{app.supp}

In the first part of this appendix, we present the initial conditions used in this work. In the second and third parts, we explain in more detail the transition probabilities $p(\ell|\ell^*)$ and $p(i,r|r^*)$ presented in the main text in Sec.~\ref{sec.timediscMar}.  Finally, in the last part, we will explain the equations that govern the dynamics of susceptible individuals with membership $k_I$.

\subsection{Initial conditions}

We assume that the fraction of people with and without access to testing are $f$ and $1-f$, respectively, and they are randomly distributed across the network nodes. Additionally, in order to reduce the number of equations that govern the dynamics between $t=0$ and $t=1$, we introduce the assumption that only a fraction $\epsilon$ of the people without access to testing services are initially infected while the rest of the population is susceptible. Thus, at $t=0$ we have the following:
\begin{itemize}
\item the total fraction of susceptible people with access to testing is $S_{\ell}(t=0)=f$, while for those without access, it is $S_{r}(t=0)=(1-f)(1-\epsilon)$,
\item the total fraction of infected people with access to testing is $I_{\ell}(t=0)=0$, while for those without access, it is $I_{r}(t=0)=\epsilon f$,
\item the fraction of susceptible people with membership $k_I$ and with access to testing is $S_{\ell}(t=0,k_I)=P(k_I)f$,
\item the fraction of susceptible people with membership $k_I$ and without access to testing is $S_{r}(t=0,k_I)=P(k_I)(1-f)(1-\epsilon)$.
\end{itemize}

On the other hand, the probability of randomly selecting a clique with ($\ell,r,i$) members at time $t=0$ is,
\begin{eqnarray}
 P_{\ell,r,i}^{t=0} = \sum_{k_C} P(k_C) \binom{k_C}{\ell, r, i}f^{\ell}[(1-\epsilon)(1-f)]^r(\epsilon f)^i  \delta_{\ell+r+i,k_C},
\end{eqnarray}
where $\delta$ is the Kronecker delta which ensures that the total number of members in the clique is equal to $k_C$.

\subsection{Transition Probability $p(\ell|\ell^*)$}\label{app.tranLead}

This section explains how to calculate $p(\ell|\ell^*)$ which is the probability of the following events occurring in an open clique with $\ell^*$ susceptible members with access to testing at time $t$: 1) exactly $\ell^*-\ell$ members will move to the $Q$ compartment during the time step $t\to t+1$, 2) $\ell$ members will remain susceptible. As discussed in the main text, Sec.~\ref{sec.timediscMar}, these two events preserve a clique's open status. Now, we will calculate the probabilities of these events below.

\subsubsection{Event 1: Remaining Susceptible}
Consider that we randomly choose an open clique “$C$” at time $t$, and then we select a susceptible member "$j$" with access to testing. This node will remain susceptible during the time step $t \to t+1$ if: 1) it is not connected to any infected nodes and 2) its neighbors (who can access testing) also have no connections to infected nodes. Mathematically, this probability is expressed as,
\begin{eqnarray}\label{eqapp.phia}
    \Phi&=&\mathcal{G}^{(1)}_{1, \ell}[\mathcal{G}^{(1)}_{1, \ell}(1,1,0),1,0].
\end{eqnarray}
To better understand this expression, we can rewrite $\Phi$ as 
\begin{eqnarray}\label{eq.phiii}
    \Phi=\mathcal{G}^{(1)}_{1, \ell}[\mathcal{G}^{(1)}_{1, \ell}(x_2,y_2,z_2),y_1,z_1],
\end{eqnarray}
with $x_2=y_2=1=y_1=1$, and $z_2=z_1=0$. In this new formulation, we have the following:
\begin{itemize}
\item $z_1=0$ indicates that "$j$" is not connected to any infected node.
\item $y_1=1$ indicates that "$j$" is connected with any number of susceptible neighbors without access to testing.
\item $\mathcal{G}^{(1)}_{1, \ell}(1,1,0)$  represents the probability of the following event. Suppose that we choose a neighbor of "$j$", and this person has regular access to testing. Then $\mathcal{G}^{(1)}_{1, \ell}(1,1,0)$ is the probability that this selected neighbor has no connections with infected people ($z_2=0$), but is connected with any number of susceptible individuals ($x_2=y_2=1$).
\end{itemize}

\subsubsection{Event 2: Entering Quarantine}
Consider again the same susceptible node "$j$" with access to testing in a randomly chosen clique “$C$” at time $t$. This node will enter quarantine during the time step $t\to t+1$ if at least one neighbor, also with access to testing, is connected to another infected node. Mathematically, the probability of this event is $\mathcal{G}^{(1)}_{1, \ell}(1,1,0) - \Phi$. Here, the first term is the probability of “$j$” " remaining susceptible or entering quarantine (but not becoming infected), while the second term is the probability that “$j$” only remains susceptible, as explained above. 

\subsubsection{Combined Transition Probability}
Considering both events, we can now derive the overall transition probability for an open clique with $\ell^*$ members with access to testing at time $t$, ending up as an open clique with $\ell\leq \ell^*$ susceptible members with access to testing:
\begin{eqnarray}
    p(\ell|\ell^*)&=& \binom{\ell^*}{\ell} \Phi^{\ell} \times (\mathcal{G}^{(1)}_{1, \ell}(1,1,0)-\Phi)^{\ell^*-\ell}.
\end{eqnarray}

\subsection{Transition Probability $p(i,r|r^*)$}\label{app.tranReg}
Here, we will calculate $p(i,r|r^*)$ which is the probability of the following events occurring in an open clique with $r^*$ susceptible members without access to testing at time $t$: 1) exactly $i$ members will contract the disease from other cliques during the time step $t\to t+1$, 2) $r$ members will remain susceptible, and 3)  $r^*-i-r$ will move to the $Q$ compartment. 

Consider a randomly chosen clique denoted by “$C$”. For each susceptible member "$j$" without access to testing, three possible transitions exist:
\begin{enumerate}
\item  “$j$” becomes infected with probability $\sigma$.
\item  “$j$” remains susceptible with probability $\Psi$.
\item  “$j$” is quarantined with probability $1-\sigma-\Psi$.
\end{enumerate}
Assuming that susceptible members without access to testing are independent of one another, we can express the transition probability $p(i,r|r^*)$ as:
\begin{eqnarray}
    p(i,r|r^*)&=&\binom{r^*}{i,r,r^*-i-r}\sigma^i \Psi^r \times (1-\sigma-\Psi)^{r^*-i-r}.
\end{eqnarray}
 In this equation, $\Psi$ and $\sigma$ are given by:
\begin{itemize}
\item  $\Psi=\mathcal{G}^{(1)}_{1, r}[\mathcal{G}^{(1)}_{1, \ell}(1,1,0),1,0]$, which has a similar interpretation to Eq. (\ref{eq.phiii}).
\item  $\sigma = G_{1,r}^t[F_{1, r}^t(0,1,1) - F_{1, r}^t(0,1,0) + F_{1, r}^t(\mathcal{G}^{(1)}_{1, \ell}(1,1,0),1,0) ]-\Psi $, which is obtained through a detailed enumeration of configurations where "$j$" becomes infected during the time step $t \to t+1$. The first term represents the probability that "$j$" either remains susceptible or becomes infected while the second term, $\Psi$, is the probability that "$j$" remains susceptible. Now, focusing on the first term, the expression:
\begin{itemize}
\item $F_{1, r}^t(0,1,1) - F_{1, r}^t(0,1,0)$ represents the probability that an open clique (in which "$j$" is a member) has at least one infected member, and no susceptible member with access to testing.
\item $F_{1, r}^t(\mathcal{G}^{(1)}_{1, \ell}(1,1,0),1,0)$ is the probability that an open clique (in which "$j$" is a member) has members with access to testing who remain disease-free and, therefore, do not initiate a quarantine order.
\end{itemize}
\end{itemize}

\subsection{Time-discrete equations for susceptible nodes with $k_I$ cliques}
Here we will derive the discrete-time equations for susceptible nodes that belong to $k_I$ cliques.  Recall that $S_{\ell}(t, k_I)$ and $S_{r}(t, k_I)$ denote the density of susceptible nodes with $k_I$ cliques at time $t$, where the subscripts $\ell$ and $r$ indicate nodes with and without access to testing, respectively.

Consider that we randomly choose at time $t$ a susceptible node without access to testing. Additionally, assume that this node has membership $k_I$. This node will remain in the susceptible compartment at $t+1$, if and only if, all its cliques stay open during the interval $t \to t+1$. Now, for a single clique to stay open, the following two conditions must be satisfied:
\begin{enumerate}
\item At time $t$, all members of the clique must be disease-free; otherwise the presence of any infected member would result in disease transmission within the clique and therefore, this clique would become closed (due to sub-step 1 of our SIQ model).
\item For those members who do have access to testing, they do not contract the infection from external sources during the interval $t \to t+1$; otherwise this clique would become closed (due to sub-step 2 of our SIQ model).
\end{enumerate}
Using the generating functions introduced in Sec.~\ref{sec.class} in the main text, we can express the probability of a single clique remaining open during the time step $t\to t+1$ as: $F_{1, r}^t\left[\mathcal{G}^{(1)}_{1, \ell}(1,1,0),1,0 \right]$. This probability consists of the following components:
\begin{enumerate}
    \item  $F_{1, r}^t(x,y,z)$ is the generating function for the probability of choosing an open clique (associated with a susceptible node without access to testing) containing $(\ell,r-1,i)$ members,
    \item  Setting $z=0$ guarantees that the clique contains no infected members.
    \item  Setting $y=1$ allows for any number of susceptible members without access to testing.
    \item $x = \mathcal{G}^{(1)}_{1, \ell}(1,1,0)$ is the probability that a susceptible member with access to testing, is connected to any number of susceptible nodes while maintaining zero connections to infected nodes.
\end{enumerate}

Then, under the assumption that the states of different cliques are statistically independent, it follows that $S_{\ell}(t+1, k_I)$ is given by, 
\begin{eqnarray}
    S_{r}(t+1, k_I) &=& S_{r}(t, k_I)\left(F_{1, r}^t\left[\mathcal{G}^{(1)}_{1, \ell}(1,1,0),1,0 \right]\right)^{k_I},
\end{eqnarray}

Following a similar reasoning, the fraction of remaining susceptible individuals with access to testing and with membership $k_I$ is given by,
\begin{eqnarray}
    S_{\ell}(t+1, k_I) &=&  S_{\ell}(t, k_I)\left(F_{1, \ell}^t\left[\mathcal{G}^{(1)}_{1, \ell}(1,1,0),1,0 \right]\right)^{k_I}.
\end{eqnarray}

\section{Additional results for $\beta=1$}\label{app.bet1}

\begin{figure}[H]
\begin{center}
\begin{overpic}[scale=0.50]{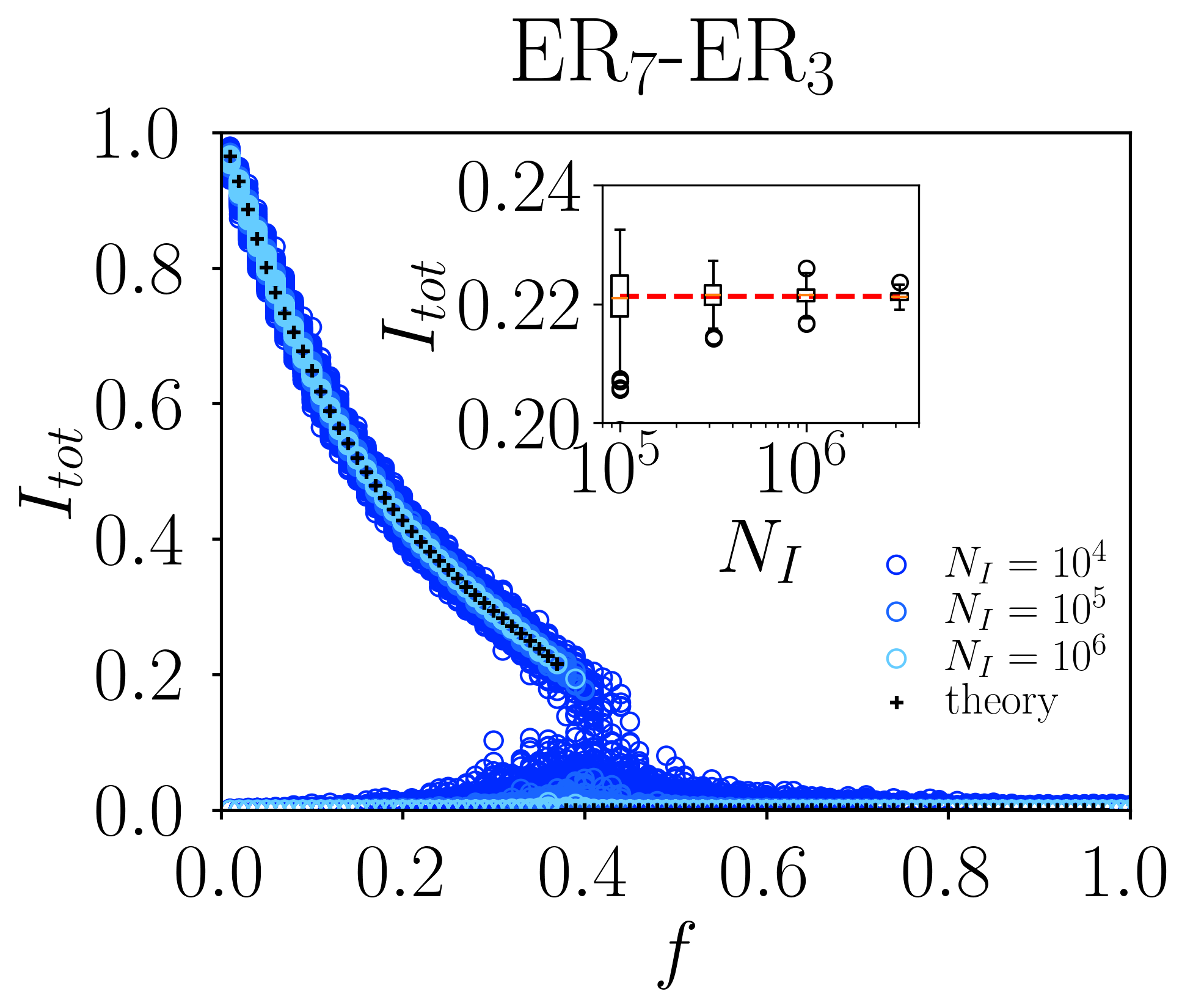}
  \put(85,28){}
\end{overpic}
\vspace{-1.1cm}
\end{center}
\caption{Fraction of the population who have ever been infected $I_{tot}$ as a function of $f$ for an ER$_7$-ER$_{3}$ network. In the main plot, the symbols "$+$" correspond to the final fraction $I_{tot}$ obtained by integrating our theoretical equations presented in Sec.~\ref{sec.class}, while the circles correspond to scatter plots obtained from stochastic simulations for 1000 realizations and for different network sizes $N_I$.  In the inset (in linear-log scale), we show box-plots for the final fraction of infected individuals $I_{tot}$ vs. $N_I$ for $f = 0.365$. The results presented here are obtained from 5000 realizations, selecting only those where $I_{tot}>0.10$ (representing roughly 5\% of the total). The horizontal dashed line marks the theoretically predicted value of $I_{tot}$ for $f = 0.365$.  For all simulations, the results correspond to the scenario in which there is only one infected individual at $t=0$, and the rest of the population is initially susceptible.}\label{fig.boxER7ER3}
\end{figure}

Here, we first demonstrate the convergence of our stochastic simulations to the theoretical predictions as the network size increases. Figure~\ref{fig.boxER7ER3} shows the fraction of infected individuals versus $f$ for ER$_7$-ER$_3$ networks with different numbers of nodes, illustrating how the simulation results progressively approach the theoretical solution with increasing network size.

Following this demonstration, we provide additional results of our SIQ model for networks with cliques in which $P(k_C)$ and $P(k_I)$ follow other degree distributions. Specifically, we considered three types of distributions:
\begin{itemize}
    \item Truncated Poisson distribution:
    \[
     Pois(\lambda,k_{min},k_{max}) = 
     \begin{cases} 
          c \frac{\lambda^k\exp(-\lambda)}{k!}, & \text{if } k_{min}\leq k \leq k_{max} \\
          0, & \text{otherwise}
     \end{cases}
    \]\label{eq.trunPoissupp}
    where $c$ is a normalization constant.
    \item Truncated Power-law distribution:
    \[
     PL(\lambda,k_{min},k_{max}) = 
     \begin{cases} 
          c k^{-\lambda}, & \text{if } k_{min}\leq k \leq k_{max} \\
          0, & \text{otherwise}
     \end{cases}
    \]\label{eq.trunPLsupp}
    where $c$ is a normalization constant.
    \item Kronecker Delta distribution:  $\delta_{k,k^*}$.
\end{itemize}

\begin{figure}[H]
\begin{center}
\begin{overpic}[scale=0.45]{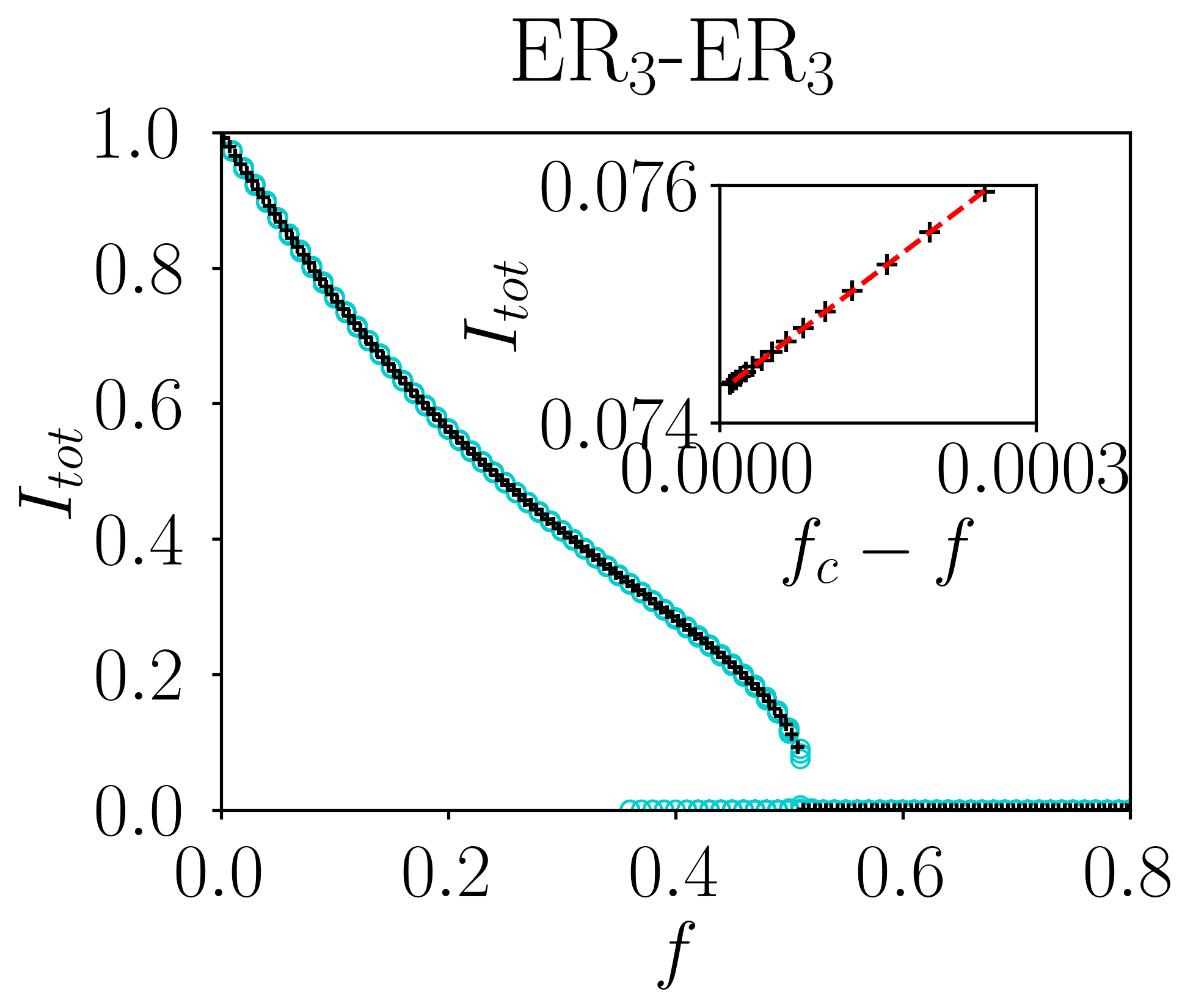}
  \put(85,28){(a)}
\end{overpic}
\begin{overpic}[scale=0.45]{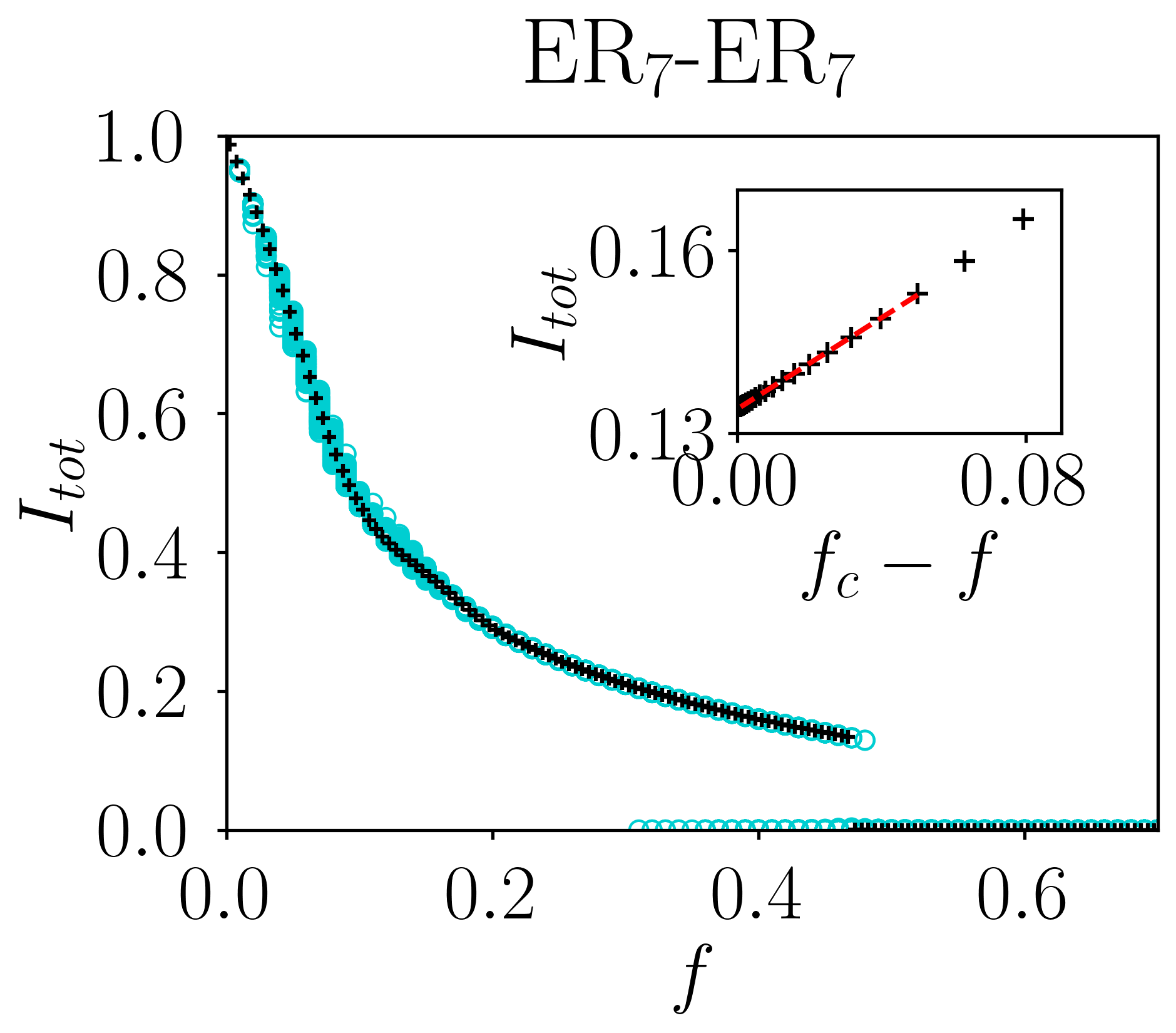}
  \put(85,28){(b)}
\end{overpic}
\begin{overpic}[scale=0.45]{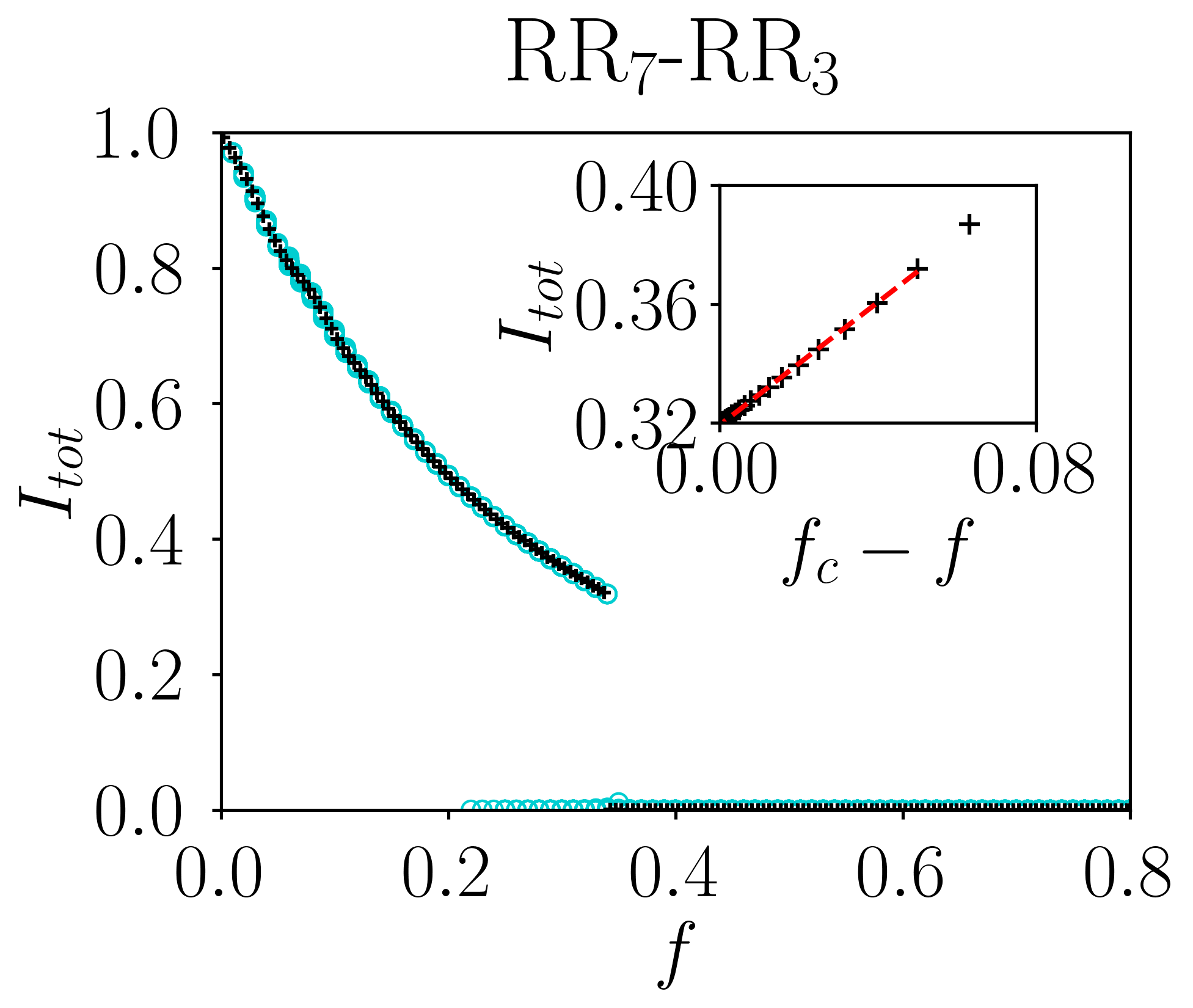}
  \put(85,28){(c)}
\end{overpic}
\begin{overpic}[scale=0.45]{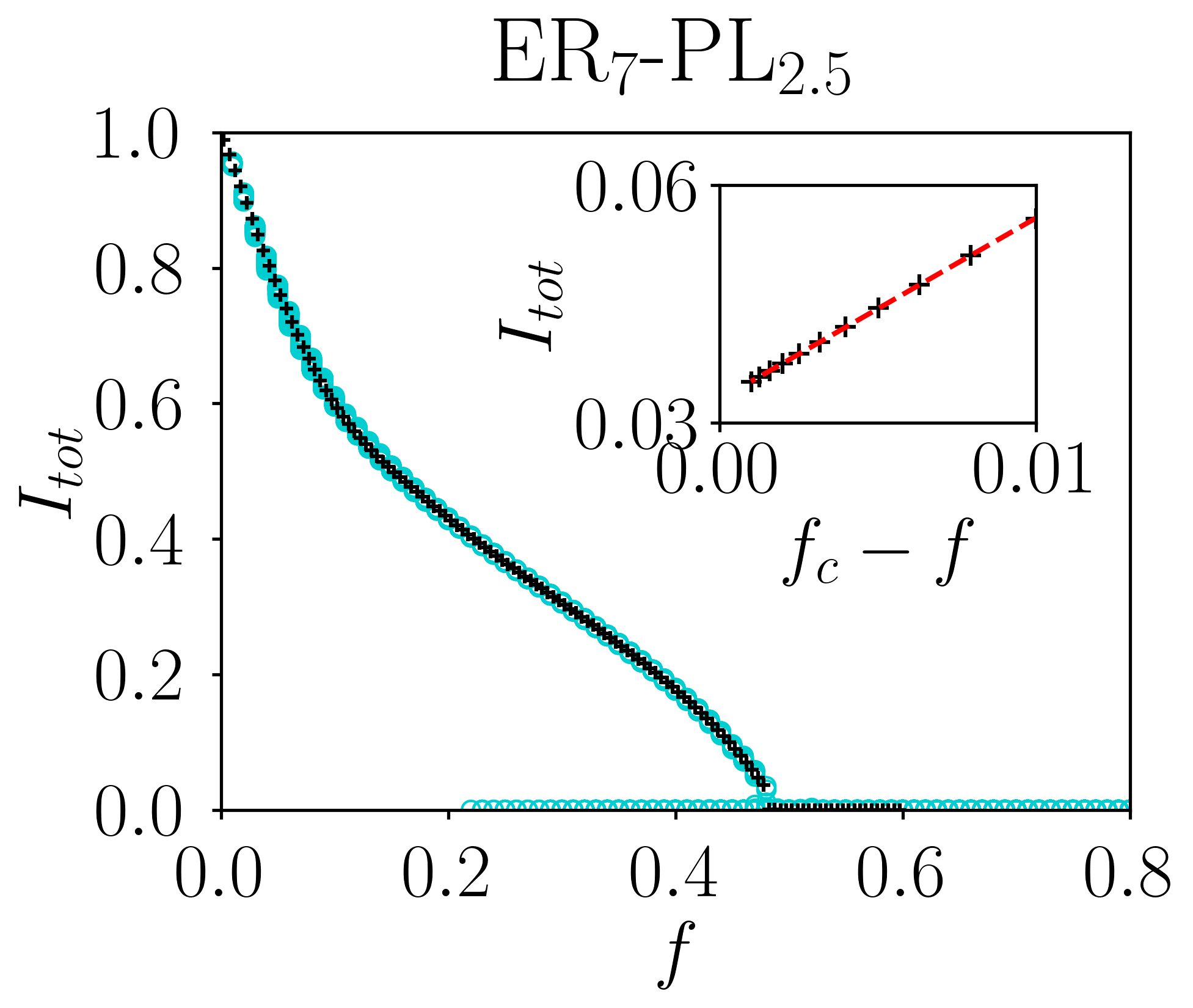}
  \put(85,28){(d)}
\end{overpic}
\vspace{-1.1cm}
\end{center}
\caption{Fraction of people who have ever been infected $I_{tot}$ as a function of $f$ for: ER$_3$-ER$_{3}$ networks (panel a), ER$_7$-ER$_{7}$ networks (panel b), RR$_7$-RR$_3$ networks (panel c), and ER$_7$-PL$_{2.5}$ networks (panel d). In the main plots, the symbols "$+$" correspond to the final fraction $I_{tot}$ obtained by integrating our theoretical equations presented in Sec.~\ref{sec.class}, while the circles correspond to a scatter plot obtained from stochastic simulations for 100 realizations on networks with $N_I=10^7$.  The inset displays $I_{tot}$ vs $f_c-f$  obtained from our theoretical equations (indicated by $+$), and a dashed line representing a linear fit.}\label{fig.suppScatt}
\end{figure}

\begin{figure}[H]
\begin{center}
\begin{overpic}[scale=0.45]{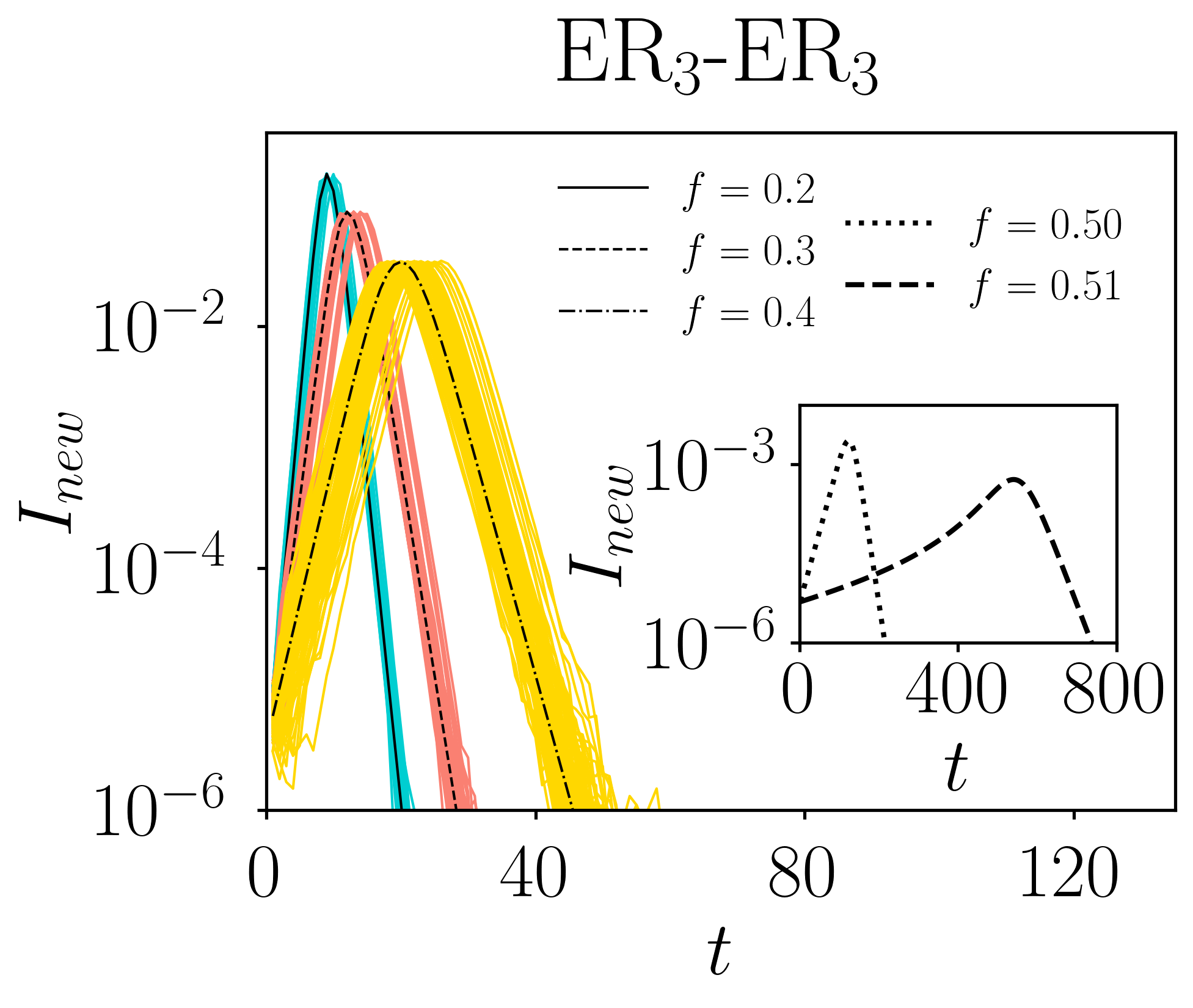}
  \put(35,65){(a)}
\end{overpic}
\begin{overpic}[scale=0.45]{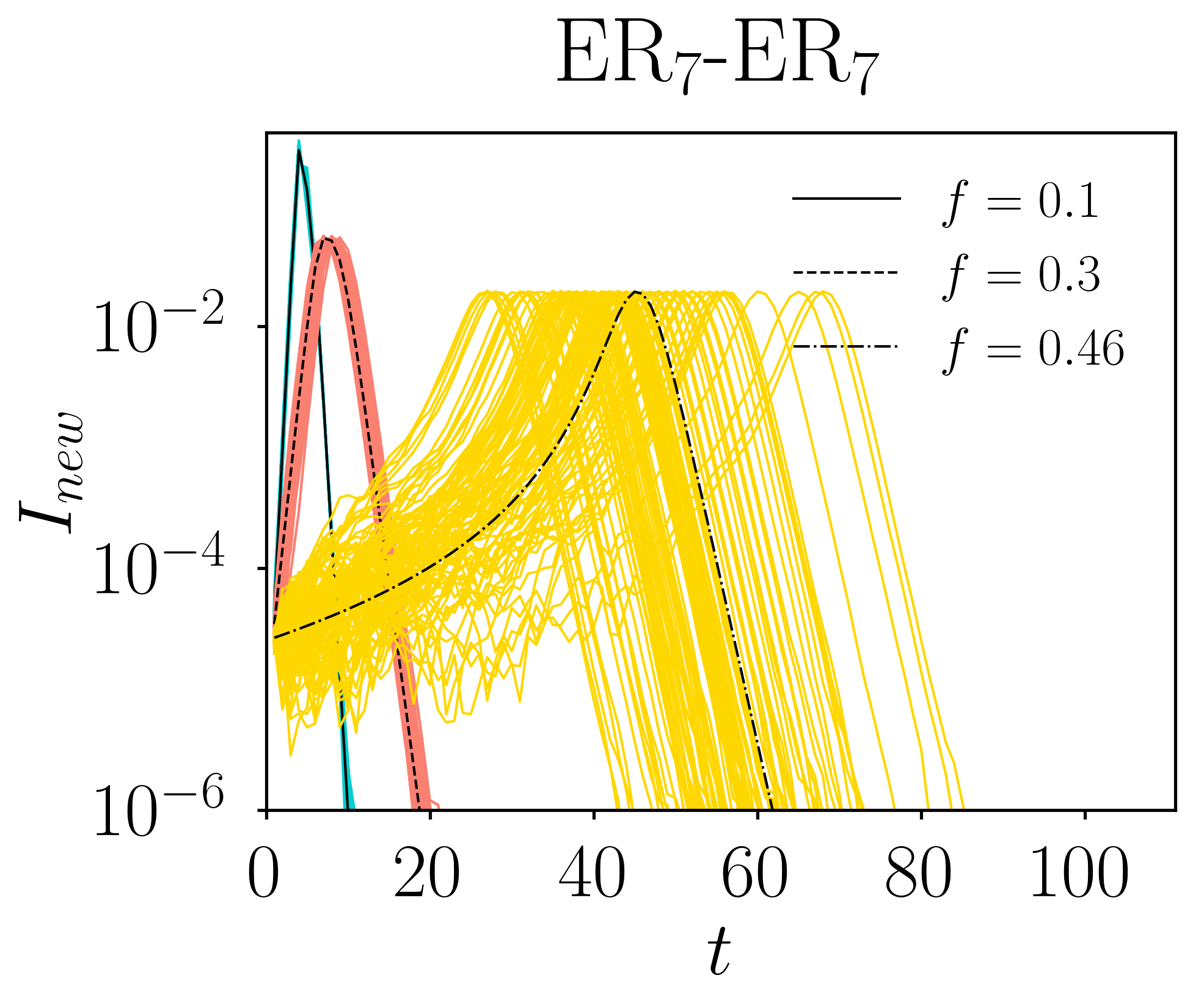}
  \put(35,65){(b)}
\end{overpic}
\begin{overpic}[scale=0.45]{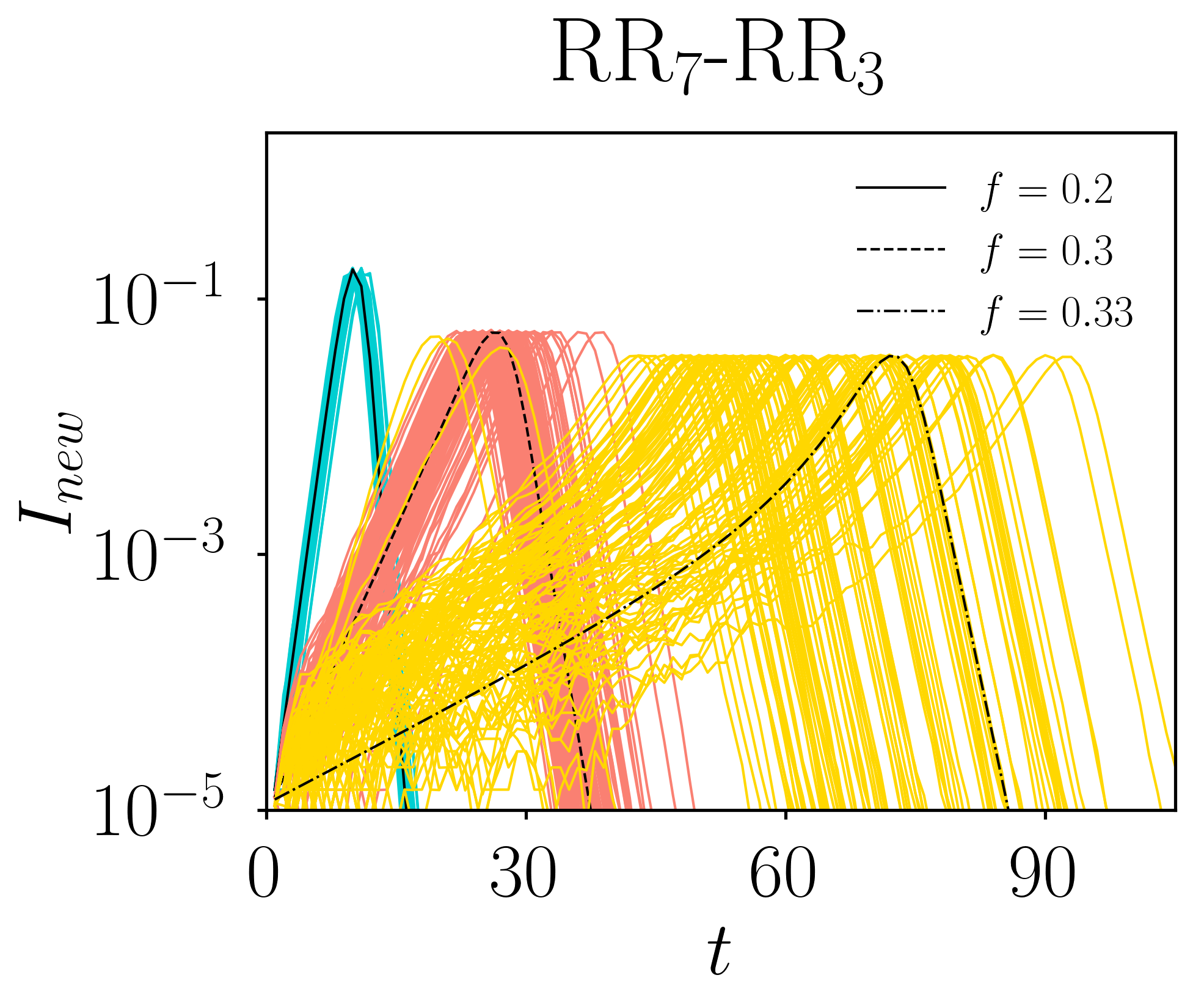}
  \put(35,65){(c)}
\end{overpic}
\begin{overpic}[scale=0.45]{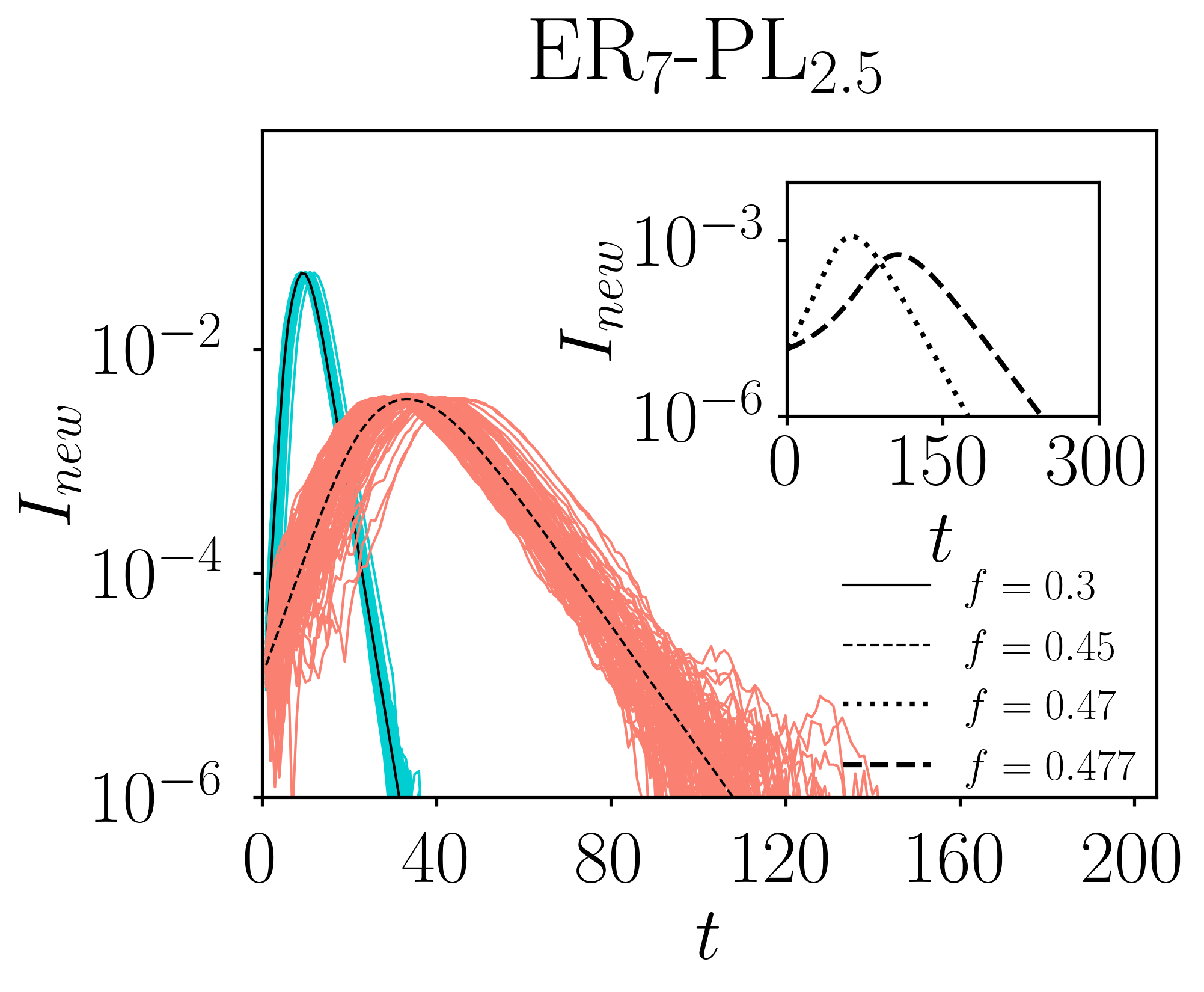}
  \put(35,65){(d)}
\end{overpic}
\vspace{-1.1cm}
\end{center}
\caption{Time evolution of $I_{new}$ for different values of $f$ in: ER$_3$-ER$_{3}$ networks (panel a), ER$_7$-ER$_{7}$ networks (panel b), RR$_7$-RR$_3$ networks (panel c), and ER$_7$-PL$_{2.5}$ networks (panel d). The vertical axis is plotted on a logarithmic scale, while the horizontal axis is linear. Black lines correspond to our theoretical solutions obtained from the equations in Sec.~\ref{sec.class} in the main text. Colored lines correspond to 100 stochastic realizations for $N_I=10^7$, except for: 1) panel b  where $N_I=4\times 10^7$ for $f=0.46$, and 2) panel d where $N_I=4\times 10^7$ for $f=0.33$. For panels a and d (ER$_3$-ER$_{3}$ and ER$_7$-PL$_{2.5}$, respectively), insets show the theoretical evolution of  $I_{new}$ for $f$ values near the critical point (in log-linear scale). These insets do not include simulation results, because finite-size effects cause significant deviations from the theoretical predictions in this region.}\label{fig.suppTime}
\end{figure}

Our results for networks with $P(k_C)\sim Pois(3,2,20)$ and $P(k_I)\sim Pois(3,1,20)$ are shown in Figs.~\ref{fig.suppScatt}a and~\ref{fig.suppTime}a. Specifically, in Fig.~\ref{fig.suppScatt}a, we display $I_{tot}$ as a function of $f$, and in Fig.~\ref{fig.suppTime}a, we show the time evolution of $I_{new}$ for different values of $f$. On the other hand, our results for networks with:
\begin{itemize}
    \item $P(k_C)\sim Pois(7,2,20)$ and $P(k_I)\sim Pois(7,1,20)$  are presented in Figs.~\ref{fig.suppScatt}b and~\ref{fig.suppTime}b.
    \item $P(k_C)\sim \delta_{k_C,7}$ and $P(k_I)\sim \delta_{k_C,3}$  are shown in Figs.~\ref{fig.suppScatt}c and~\ref{fig.suppTime}c.
    \item $P(k_C)\sim Pois(7,2,20)$ and $P(k_I)\sim PL(2.5,2,50)$  are presented in Figs.~\ref{fig.suppScatt}d and~\ref{fig.suppTime}d. For the distribution $P(k_I)$, we chose $k_{min}=2$, because for $k_{min}=1$ we do not observe the super-exponential growth phenomenon.
\end{itemize}

\begin{figure}[H]
\begin{center}
\begin{overpic}[scale=0.4]{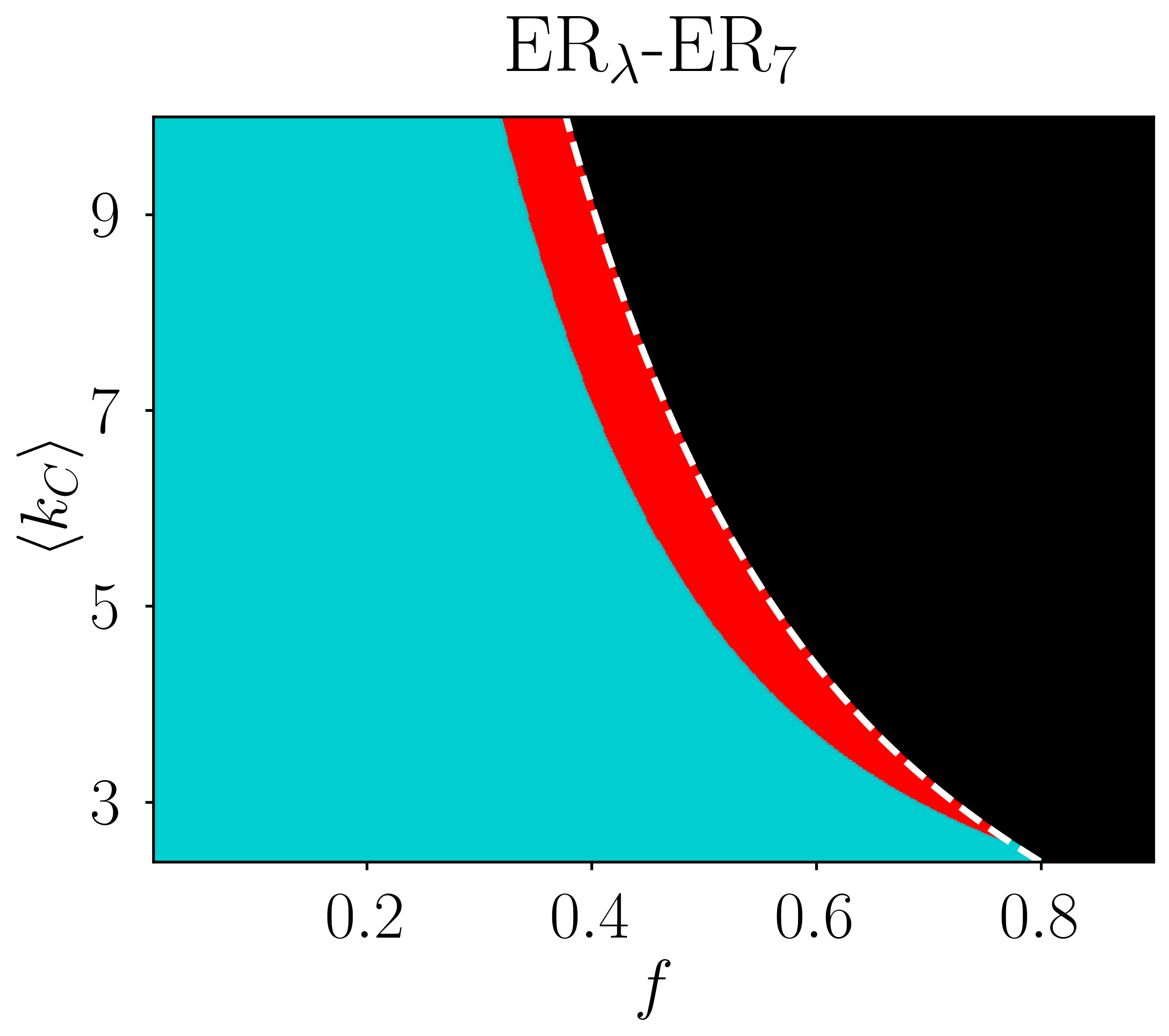}
  \put(25,28){(a)}
\end{overpic}
\begin{overpic}[scale=0.4]{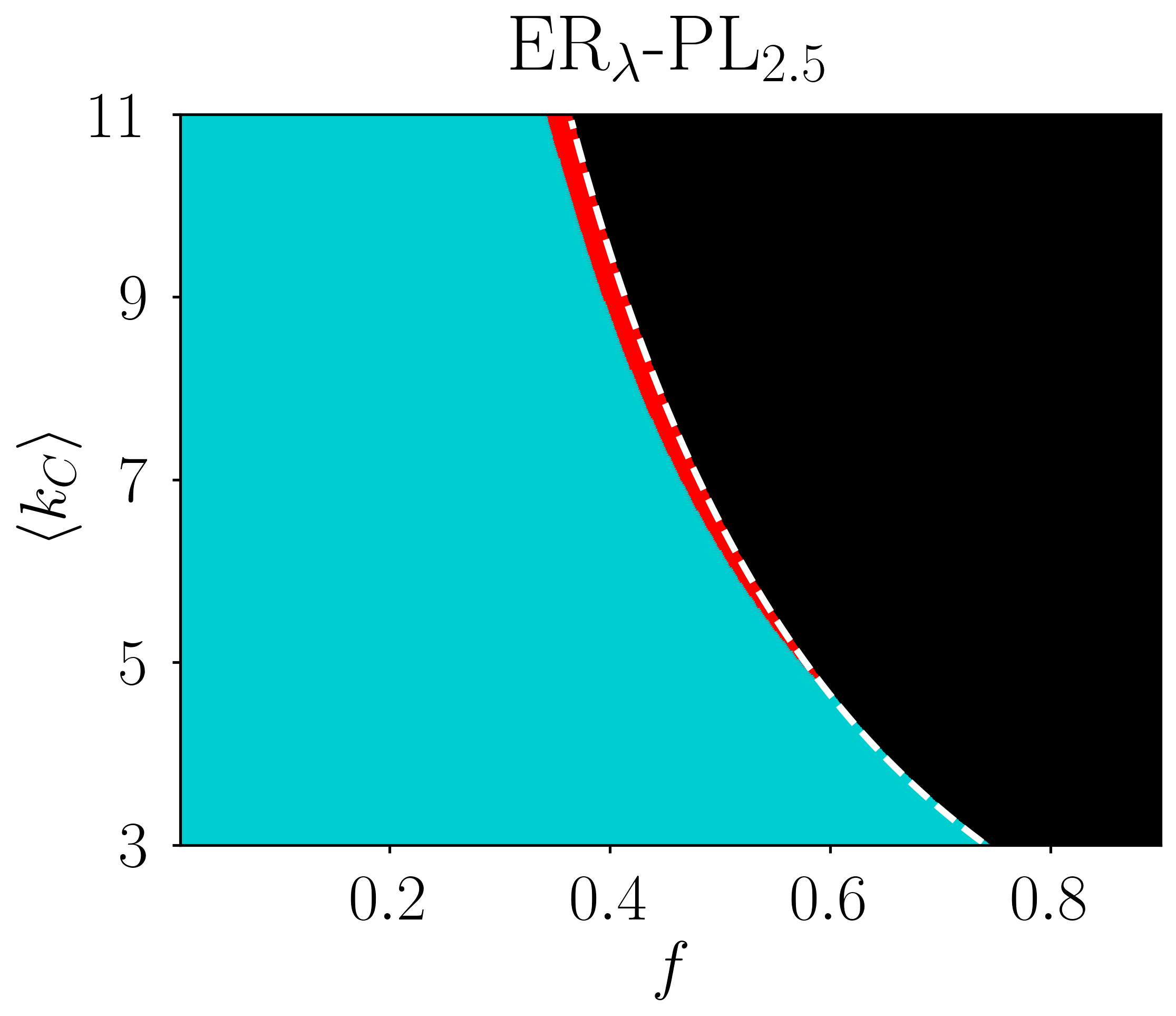}
  \put(25,28){(b)}
\end{overpic}
\vspace{-1.1cm}
\end{center}
\caption{Phase diagram in the $\langle k_C\rangle$-$f$ plane, where clique sizes follow a truncated Poisson distribution Pois($\lambda$,2,20) with $\lambda \in (1,11)$, and the membership $k_I$ follows a: 1) truncated Poisson distribution Pois(7,2,20) (panel a), and 2) truncated power-law distribution PL(2.5,2,50) (panel b). In this figure, $\langle k_C\rangle$ represents the average clique size. The light blue region corresponds to the area in the plane where the super-exponential growth dynamics is not detected. The red region indicates super-exponential growth, while the black region is free of epidemics. The white dashed line represents the threshold where the basic reproduction number equals one (see Eq.(\ref{eq.r00})).}\label{fig.suppHeat}
\end{figure}

Additionally, in Fig.~\ref{fig.suppHeat}, we display the phase diagram of our model in the  $\langle k_C\rangle-f$ plane for networks with:
\begin{itemize}
    \item $P(k_C)\sim Pois(\lambda,2,20)$  and $P(k_I)\sim Pois(7,1,20)$,
    \item $P(k_C)\sim Pois(\lambda,2,20)$  and $P(k_I)\sim PL(2.5,2,50)$,
\end{itemize}
where $\langle k_C \rangle =\sum_{k_C=2}^{k_C=20} c k_C  \frac{\lambda^{k_C}\exp(-\lambda)}{k_C!}$.

We note that all the results shown in Figs.~\ref{fig.suppScatt}-\ref{fig.suppHeat} are qualitatively consistent with those presented in the main text. In addition, in Appendix~\ref{app.low1}, we also show that the discontinuous transition and the super-exponential growth phenomenon are also observed for $\beta<1$.

\section{Numerical results for $\beta<1$}\label{app.low1}
\begin{figure}[H]
\begin{center}
\begin{overpic}[scale=0.40]{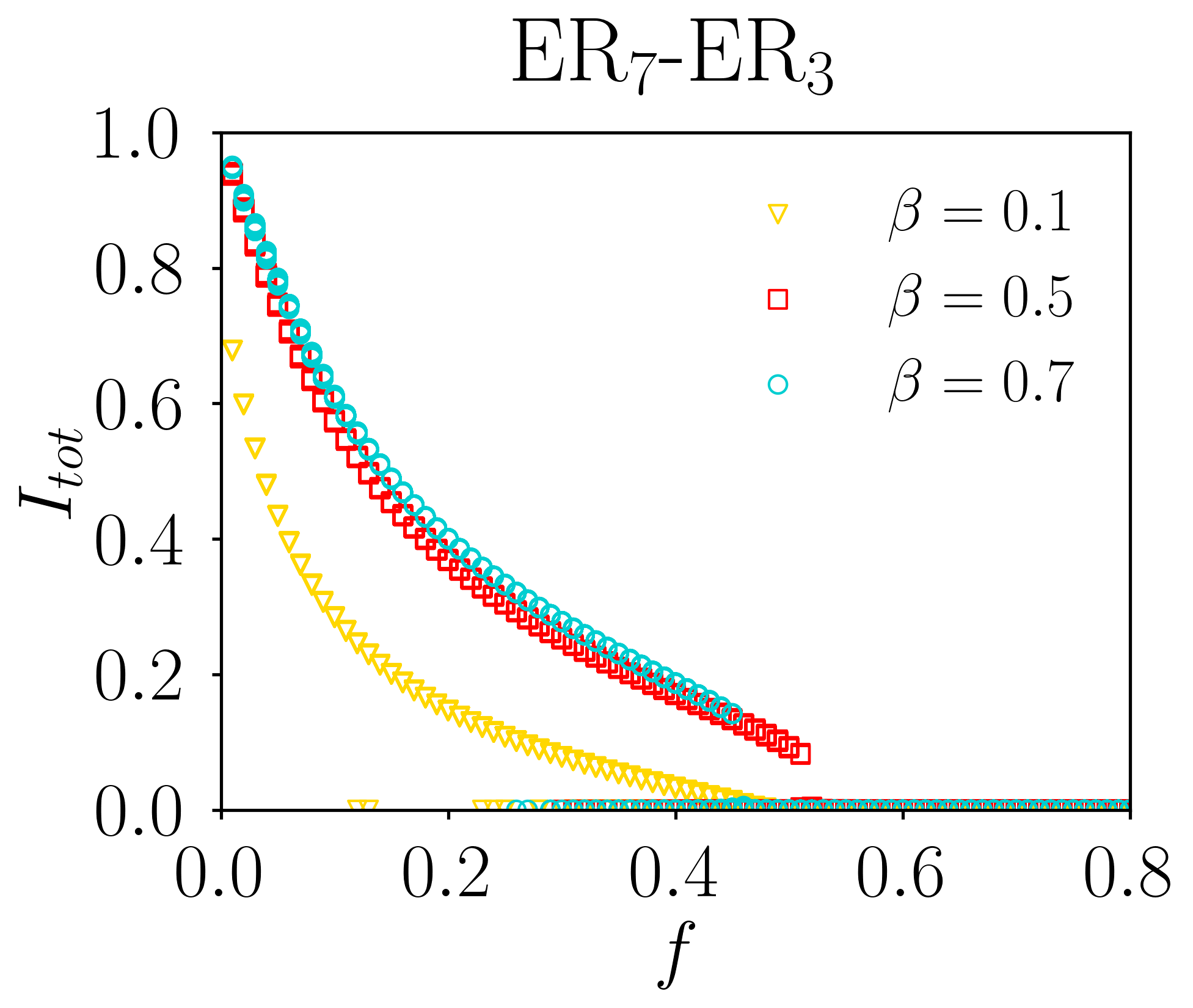}
  \put(85,28){(a)}
\end{overpic}
\begin{overpic}[scale=0.40]{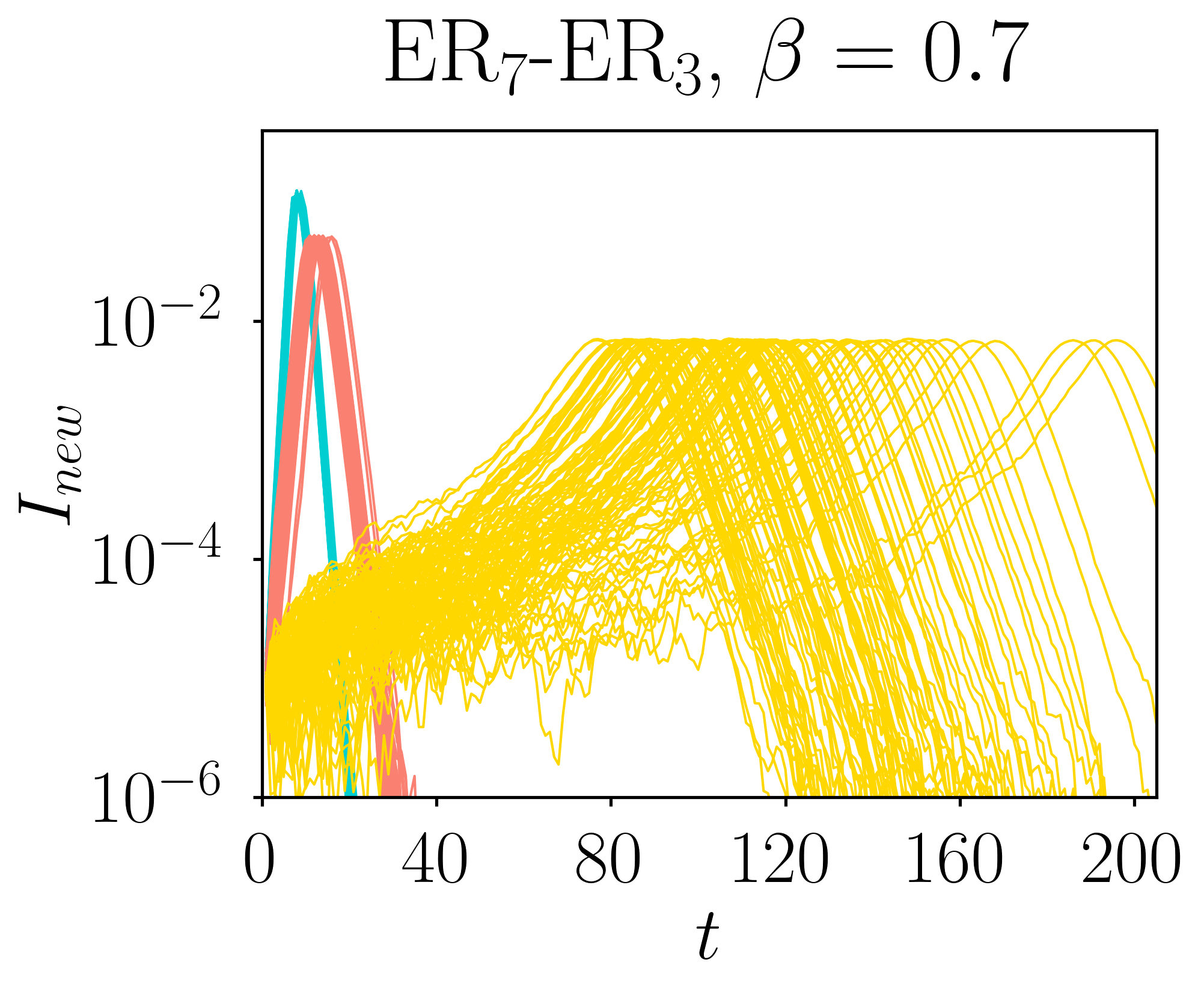}
  \put(85,28){(b)}
\end{overpic}
\begin{overpic}[scale=0.40]{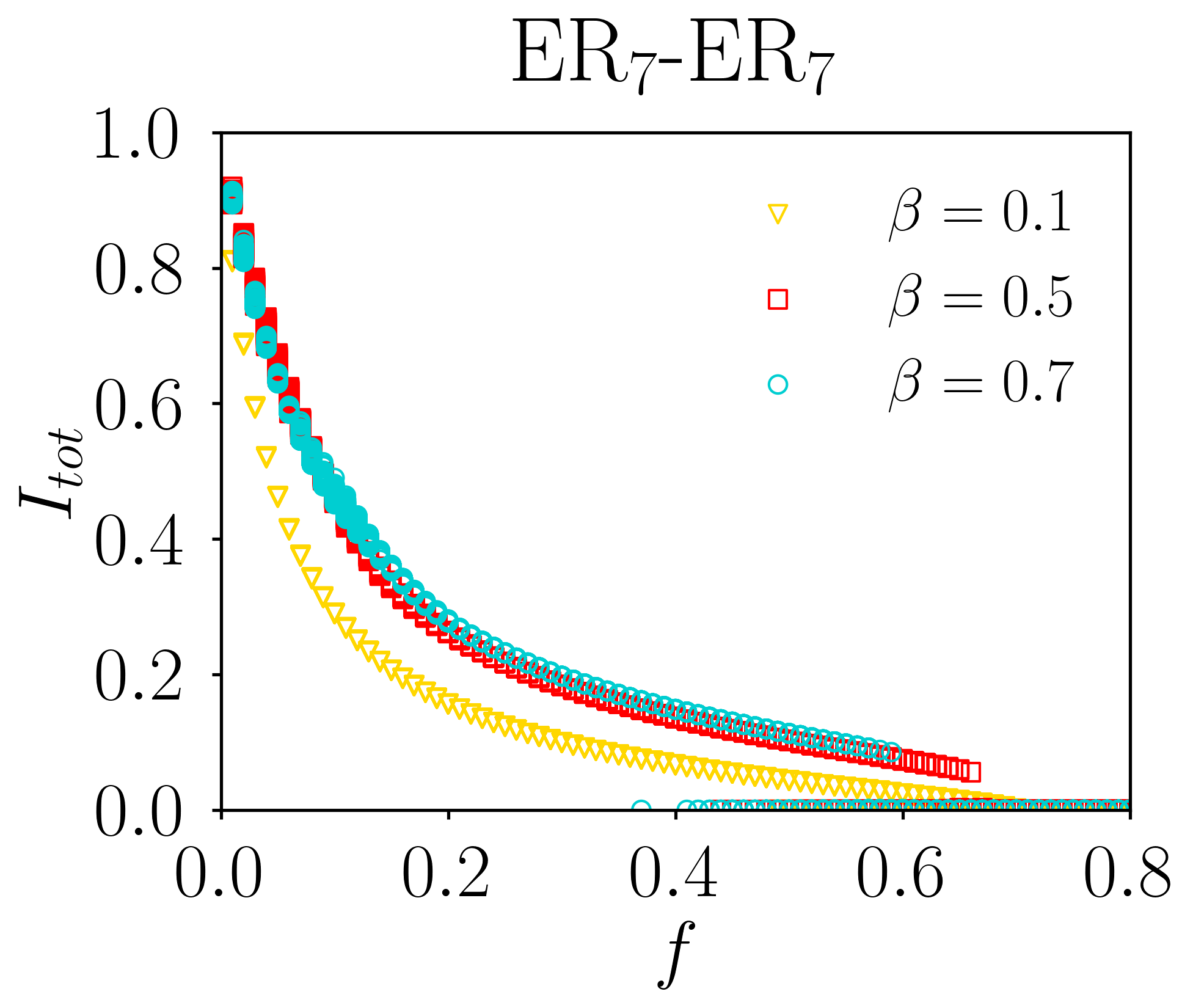}
  \put(85,28){(c)}
\end{overpic}
\begin{overpic}[scale=0.40]{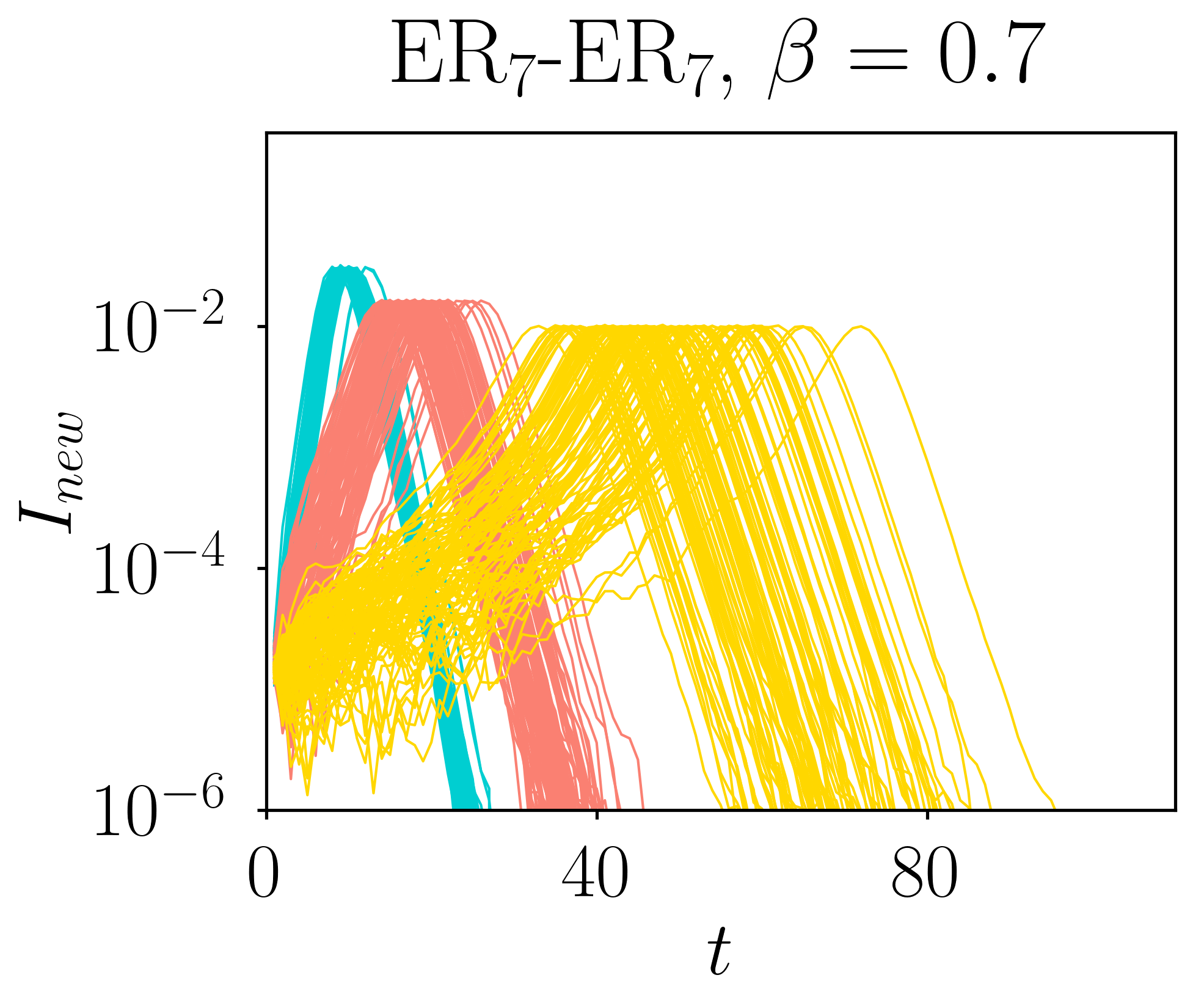}
  \put(85,28){(d)}
\end{overpic}
\begin{overpic}[scale=0.40]{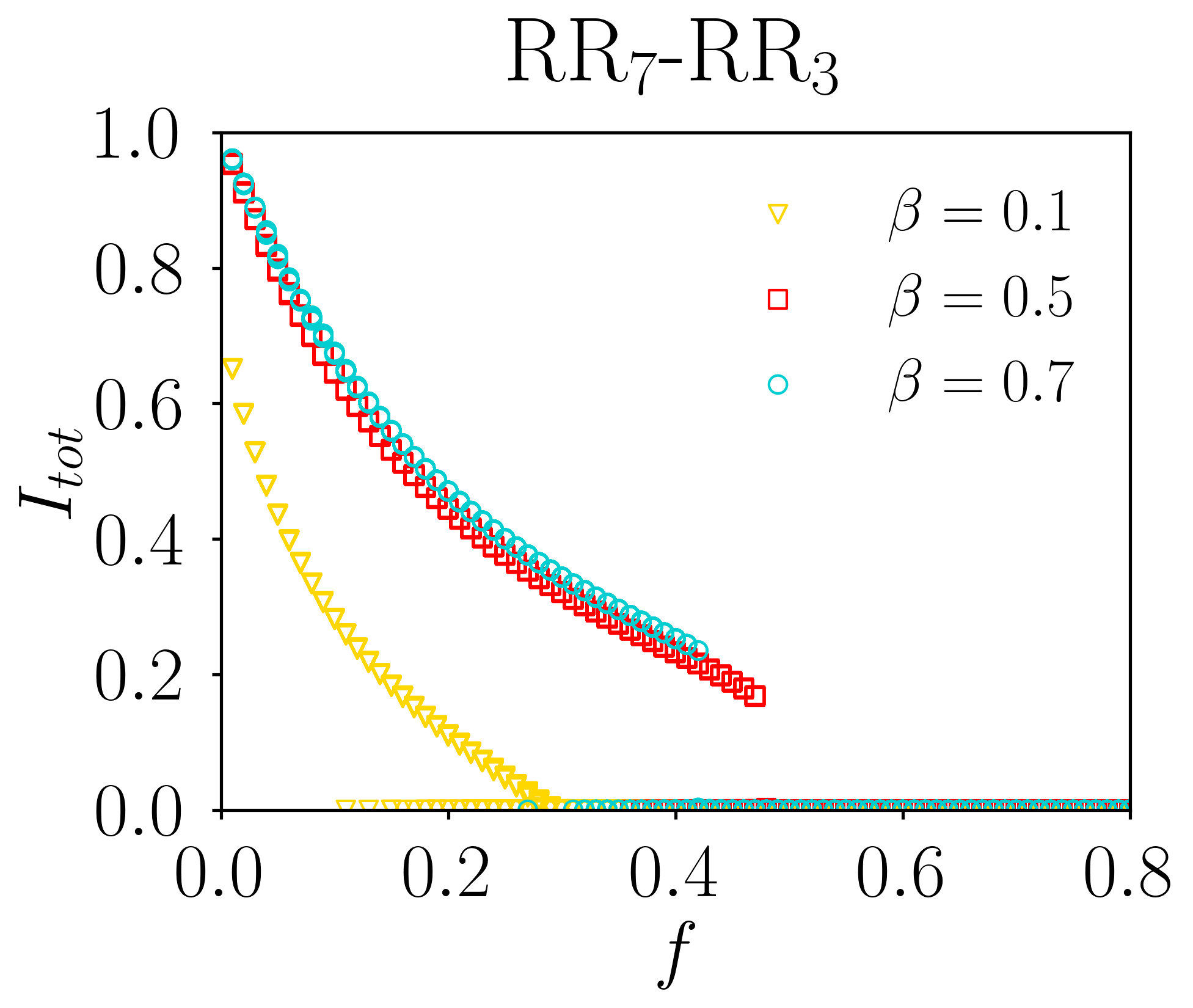}
  \put(85,28){(e)}
\end{overpic}
\begin{overpic}[scale=0.40]{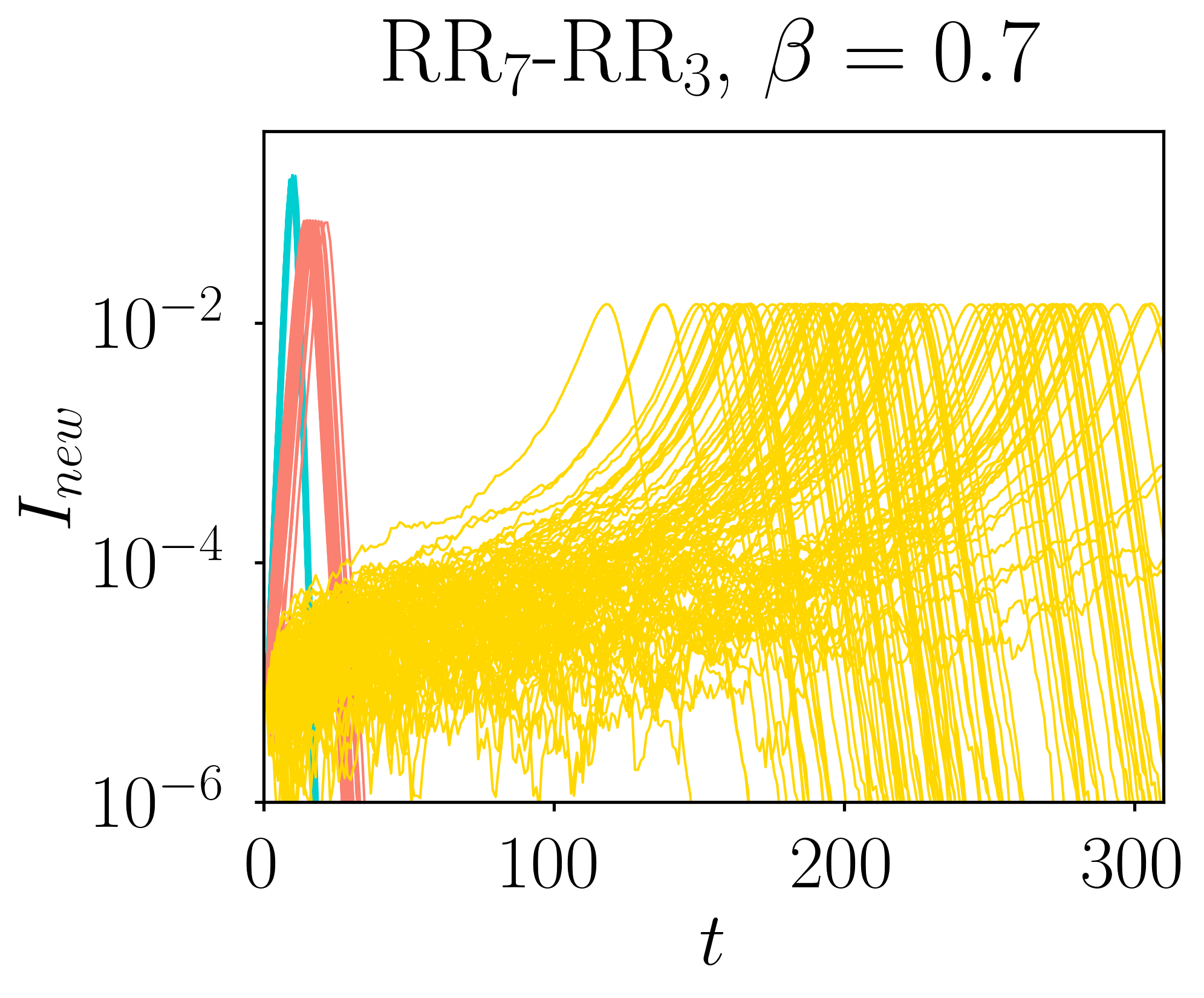}
  \put(85,28){(f)}
\end{overpic}
\vspace{-1.1cm}
\end{center}
\caption{Numerical results for different network topologies and $\beta<1$. Left column: Scatter-plots of $I_{tot}$ against $f$ for different values of $\beta$ ($\beta=0.7$, $\beta=0.5$, and $\beta=0.1$) in a: (a) ER$_7$-ER$_3$ network, (c) ER$_7$-ER$_7$ network, and (e) RR$_7$-RR$_3$ network, all with $N_I=10^7$. Numerical results were obtained from 100 network realizations. Right column: 100 simulation trajectories of the number of new cases for $\beta=0.7$ with various $f$ values: (b) ER$_7$-ER$_3$ network with $f=0.2$ (light blue, $N_I=10^7$), $f=0.3$ (red, $N_I=10^7$), and $f=0.45$ (yellow, $N_I=4\times 10^7$); (d) ER$_7$-ER$_7$ network with $f=0.40$ (light blue, $N_I=10^7$), $f=0.50$ (red, $N_I=10^7$), and $f=0.56$ (yellow, $N_I=4\times 10^7$); (f) RR$_7$-RR$_3$ network with $f=0.2$ (light blue, $N_I=10^7$), $f=0.3$ (red, $N_I=10^7$), and $f=0.42$ (yellow, $N_I=4\times 10^7$).}\label{fig.suppbetlow}
\end{figure}

In this appendix, we present our numerical results of the SIQ model for $\beta<1$ across different network topologies. In Figs.~\ref{fig.suppbetlow}a, c, and e, we show $I_{tot}$ at the final stage as a function of $f$ for several values of $\beta$, and from these figures we can observe that for higher values of $\beta$, our model exhibits an abrupt transition as in the case of $\beta=1$ shown in the main text. On the other hand, in Figs.~b, d, and f, we display the time evolution of the number of new cases $I_{new}$ for $\beta=0.7$ and several values of $f$. Our results suggest that well below the transition point, $I_{new}$ increases as an exponential function, whereas for $f\lesssim f_c$, the number of new cases grows faster than an exponential function.

\bibliography{bib}

\end{document}